\documentclass[12pt, runningheads]{llncs}
\usepackage{hyperref}
\usepackage{graphicx}
\usepackage{tikz-cd}
\usepackage{amsmath}
\usepackage{amssymb}
\usepackage{thm-restate}
\usepackage[capitalise]{cleveref}
\usepackage{fullpage}
\usepackage{algorithm}
\usepackage{algpseudocode}
\usepackage[utf8]{inputenc}
\usepackage{subcaption}

\usetikzlibrary{patterns}
\usetikzlibrary{decorations.pathreplacing,calligraphy}

\newlength{\hatchspread}
\newlength{\hatchthickness}
\newlength{\hatchshift}
\newcommand{\hatchcolor}{}
\tikzset{hatchspread/.code={\setlength{\hatchspread}{#1}},
         hatchthickness/.code={\setlength{\hatchthickness}{#1}},
         hatchshift/.code={\setlength{\hatchshift}{#1}},
         hatchcolor/.code={\renewcommand{\hatchcolor}{#1}}}
\tikzset{hatchspread=3pt,
         hatchthickness=0.4pt,
         hatchshift=0pt,
         hatchcolor=black}
\pgfdeclarepatternformonly[\hatchspread,\hatchthickness,\hatchshift,\hatchcolor]
   {custom north west lines}
   {\pgfqpoint{\dimexpr-2\hatchthickness}{\dimexpr-2\hatchthickness}}
   {\pgfqpoint{\dimexpr\hatchspread+2\hatchthickness}{\dimexpr\hatchspread+2\hatchthickness}}
   {\pgfqpoint{\dimexpr\hatchspread}{\dimexpr\hatchspread}}
   {
    \pgfsetlinewidth{\hatchthickness}
    \pgfpathmoveto{\pgfqpoint{0pt}{\dimexpr\hatchspread+\hatchshift}}
    \pgfpathlineto{\pgfqpoint{\dimexpr\hatchspread+0.15pt+\hatchshift}{-0.15pt}}
    \ifdim \hatchshift > 0pt
      \pgfpathmoveto{\pgfqpoint{0pt}{\hatchshift}}
      \pgfpathlineto{\pgfqpoint{\dimexpr0.15pt+\hatchshift}{-0.15pt}}
    \fi
    \pgfsetstrokecolor{\hatchcolor}
    \pgfusepath{stroke}
   }

\usepackage{url}
\tikzset{myfill/.style ={fill=blue!20!white, draw=black}}
\tikzset{myfill_noborder/.style ={fill=blue!20!white, draw=blue!20!white}}

\tikzset{myredfill/.style ={red!20!white, draw=black, dashed}}
\tikzset{myredfillnotdashed/.style ={red!20!white, draw=black}}
\tikzset{mygreenfill/.style ={green!20!white, draw=black}}
\tikzset{myyellowfill/.style ={yellow!20!white, draw=black, dashed}}

\tikzset{bart_yellow/.style ={yellow!25!white, draw=none}}
\tikzset{bart_green/.style ={green!20!white, draw=none}}
\tikzset{bart_orange/.style ={orange!20!white, draw=none}}
\tikzset{bart_purple/.style ={purple!25!white, draw=none}}

\def\strip#1#2{\draw[thick] (0, #2) -- (0,0) -- (#1,0) -- (#1,#2);
               \draw[thick,->] ( 0,#2) -> +(0,0.5);
               \draw[thick,->] (#1,#2) -> +(0,0.5);
               \foreach \x in {0,...,#1}
                  \draw (\x cm, 2pt) -- (\x cm, -2pt) node[anchor=north] {\footnotesize$\x$};
              \foreach \y in {0,...,#2}
                  \draw (2pt,\y cm) -- (-2pt,\y cm) node[anchor=east] {\footnotesize$\y$};}

\def\hbl{h_{\text{BL}}}
\def\hopt{h_{\text{OPT}}}
\def\hblbest{h_{\text{BL}}^{\text{best}}}
\def\hblworst{h_{\text{BL}}^{\text{worst}}}
\def\hkbl{h_{\text{$k$-BL}}}
\newcommand{\R}{\mathbb{R}}

\spnewtheorem{construction}{Construction}{\bfseries}{\normalfont}

\begin{document}
\title{The Bottom-Left Algorithm for the Strip~Packing~Problem}
\author{Stefan Hougardy\thanks{ funded by the Deutsche Forschungsgemeinschaft (DFG, German Research Foundation) under Germany's Excellence Strategy -- EXC-2047/1 -- 390685813}\inst{1}
\and Bart Zondervan\inst{2}}

\authorrunning{S.\ Hougardy and B.\ Zondervan}
%

\institute{Research Institute for Discrete Mathematics and Hausdorff Center for Mathematics, University of Bonn, Germany, \email{hougardy@dm.uni-bonn.de} \and
Faculty of Mathematics and Computer Science, University of Bremen, Germany, \email{bart.zondervan@uni-bremen.de}
}
\maketitle              
\begin{abstract}
    The bottom-left algorithm is a simple heuristic for the Strip Packing Problem.
    It places the rectangles in the given order at the lowest free position in the strip, 
    using  the left most position in case of ties.  
    Despite its simplicity, the exact approximation ratio of the bottom-left algorithm remains unknown. 
    We will improve the more-than-40-year-old value for the lower bound from $5/4$ to $4/3 - \varepsilon$. 
    Additionally, we will show that this lower bound holds even in the special case of squares, where the previously known lower bound was $12/11 -\varepsilon$. These lower bounds apply regardless of the ordering of the rectangles.  
    When squares are arranged in the worst possible order, we establish a constant upper bound and a $10/3-\varepsilon$  lower bound for the approximation ratio of the bottom-left algorithm.
    This bound also applies to some online setting and yields an almost tight result there.
    Finally, we show that the approximation ratio of a local search algorithm based on permuting rectangles in the ordering of the bottom-left algorithm is at least~$2$ and that such an algorithm may need an exponential number of improvement steps to reach a local optimum.

\keywords{Bottom-Left Algorithm \and Strip Packing \and Approximation Algorithm.}
\end{abstract}

\section{Introduction}
In the Strip Packing Problem, a rectangular strip of fixed width and infinite height is given.
The task is to find an orthogonal packing of a given set of rectangles into the strip such that no two rectangles overlap and the total height of the packing is minimal.
Rotation of rectangles is not allowed.

A reduction from the Bin Packing Problem shows that Strip Packing is NP-hard \cite{Karp}. 
It is even strongly NP-hard \cite{Garey78}.
Moreover, this reduction establishes that unless P=NP, there cannot exist a $(3/2-\varepsilon)$-approximation algorithm for Strip Packing.
Currently, the best-known approximation algorithm  achieves an approximation ratio of $5/3+\varepsilon$ \cite{harren2014,Galvez2023}.
However, this algorithm is rather complicated and may not be of practical relevance.

In contrast, the bottom-left algorithm is extremely simple. It operates by packing the rectangles in the given order, positioning them at the lowest available point within the strip.
In situations where there are multiple lowest positions possible, the bottom-left algorithm selects the left-most
of these positions.
An implementation of the bottom-left algorithm with quadratic time complexity is presented in~\cite{Chazelle1983}.

The approximation ratio of the bottom-left algorithm heavily depends on the ordering of the rectangles.
It is easy to construct instances and orderings of the rectangles such that the approximation ratio
of the bottom-left algorithm is arbitrarily bad. Baker, Coffman, and Rivest~\cite{baker:1980}
have shown that this may even happen if the rectangles are ordered by increasing width.
Contrary to this they proved~\cite{baker:1980} that when the rectangles are ordered by decreasing 
width then the bottom-left algorithm has approximation ratio~3. In case that all rectangles are squares they
proved an approximation ratio of~2 for the bottom-left algorithm. 

A natural question arising in this context is: \emph{What is the approximation ratio of the bottom-left algorithm
if we have a best possible ordering of the input rectangles?} It is tempting to expect that among 
the $n!$ possible orderings of the given $n$ rectangles there is always one such that the
bottom-left algorithm will find an optimum solution. However, this is not the case as was shown in 1980
by Brown~\cite{brown:1980}: There exist instances 
of the strip packing problem for which the bottom-left algorithm cannot achieve an approximation ratio
better than $5/4$ not even for the best ordering of the rectangles. In case of squares it was shown
by Baker, Coffman, and Rivest~\cite{baker:1980} that for any fixed $\varepsilon > 0$ the bottom-left algorithm 
cannot achieve an approximation ratio of $12/11-\varepsilon$.
Thus there remain large gaps between $5/4$ and $3$ for the approximation ratio of the bottom-left 
algorithm for the Strip Packing Problem in case of rectangles and between~$12/11-\varepsilon$ and~$2$ 
in case of squares.
We will narrow these gaps by improving the lower bound in both cases to $4/3-\varepsilon$.
This is the first improvement on these bounds after more than 40 years.

\begin{restatable}{theorem}{thmlowerboundsquares}
\label{thm:best_BL_square_lower_bound}
    For all $\varepsilon>0$ the approximation ratio of the bottom-left algorithm for the Square Strip Packing Problem cannot be better than $4/3-\varepsilon$ even if the squares are ordered in the best possible way.     
\end{restatable}

Additionally, instead of looking at the best ordering, we will also look at the worst ordering.
As mentioned above the approximation ratio of the bottom-left algorithm might be unbounded when 
the rectangles are badly ordered.
On the contrary, we can show that for squares the approximation ratio of the bottom-left algorithm is always bounded, regardless of the ordering.
We also construct a $10/3-\varepsilon$ lower bound for this case improving 
the so far best lower bound of~$2-\varepsilon$~\cite{baker:1980}.

\begin{restatable}{theorem}{thmworstcaseordering}
\label{thm:worst_case_ordering_bounds}
   The bottom-left algorithm has constant approximation ratio for the Square Strip Packing Problem, 
   for all possible orderings of the squares. 
   This approximation ratio cannot be better than $10/3-\varepsilon$. 
\end{restatable}

Our lower bound also applies to an online version of the bottom-left algorithm that was studied in~\cite{fekete2014online} and was shown to have approximation ratio $3.5$. We therefore get
an almost tight result for this case:

\begin{restatable}{corollary}{corlowerboundfekete}
\label{cor:lower-bound-fekete}
   The online \textit{BottomLeft} algorithm from~\cite{fekete2014online} has approximation ratio between $10/3$ and $3.5$. 
\end{restatable}

Last of all, we study a local search variant of the bottom-left algorithm.
The bottom-left $k$-local search algorithm starts with an initial bottom-left packing and in each iteration the algorithm tries to permute $k$ rectangles such that the bottom-left algorithm on the new ordering returns a packing with strictly lower height.
Firstly, we show a lower bound equal to $2$ for the bottom-left $k$-local search algorithm, implying that this algorithm cannot find an ordering such that the bottom-left algorithm has approximation ratio better than the currently best-known $(5/3+\varepsilon)$-approximation ratio from~\cite{harren2014}.
Secondly, we also show that the local search algorithm may need an exponential number of 
iterations before reaching a local optimum.

\begin{restatable}{theorem}{thmlocalsearch}
\label{thm:local_search}
   The approximation ratio of the $k$-local search bottom-left algorithm is bounded from below by~2, even in case of squares. Moreover, this algorithm may need an exponential number of iterations to find a local optimum. 
\end{restatable}

\paragraph{Outline of the paper}
After starting with some basic definitions in~\cref{sec:preliminaries} we will 
present in~\cref{section:improved_lower_bound} our new lower bounds for the 
bottom-left algorithm assuming a best possible ordering of the rectangles. 
We first consider the general Strip Packing case, and afterwards show that 
the lower bound $4/3-\varepsilon$ also holds in the Square Strip Packing case.
Next,~\cref{section:worst_case_BL} shows that in case of squares 
the approximation ratio of the bottom-left algorithm is bounded by a constant, even for the worst ordering of the squares. 
We also prove a lower bound of $10/3-\varepsilon$ for this case. 
Last of all,~\cref{section:local_search} studies the bottom-left $k$-local search algorithm.
We prove that this novel local search algorithm has approximation ratio no better than~$2$.
Moreover, we show that this algorithm might take an exponential number of improvement steps to reach a local optimum.

\section{Preliminaries}
\label{sec:preliminaries}
A Strip Packing instance $\cal I$ consists of a vertical strip  of fixed width $W$ and infinite height 
together with a set $R=\{r_1, \ldots, r_n\}$ of $n$ closed rectangles. 
Each rectangle $r_i$ has a given height $h_i := h(r_i)$ and width~$w_i :=~w(r_i)$. 
Assume that $\max\{w_i : 1\le i \le n\} \le W$.
A packing of $R$ into the strip is defined by specifying the lower left coordinate $(x_i, y_i)$ for each $r_i \in R$. 
A packing of $R$ is \emph{feasible} if all rectangles lie within the strip and no two rectangles overlap within their interior, i.e., the following two conditions have to be satisfied:
\begin{eqnarray*}
   x_i \ge 0, ~x_i + w_i \le W, ~y_i \ge 0 &\mbox{ for all $1\le i\le n$,} \\
   (x_i, x_i + w_i) \times (y_i, y_i + h_i) \cap 
   (x_j, x_j + w_j) \times (y_j, y_j + h_j) = \emptyset &\mbox{ for all $1\le i < j \le n$.}
\end{eqnarray*}

The height of a feasible packing is the maximum of $\{y_i + h_i : 1\le i \le n\}$.
The goal of the Strip Packing Problem is to compute a feasible packing of minimum height for a given 
Strip Packing Instance $\cal I$. We denote this value by $\hopt({\cal I})$. Note that our definition
of the Strip Packing Problem does not allow to rotate the rectangles by~90~degrees.
The Square Strip Packing Problem is the special case of the Strip Packing Problem where all rectangles are squares.
We will abbreviate the Strip Packing Problem respectively the Square Strip Packing Problem by SPP respectively SSPP.

Given a Strip Packing Instance $\mathcal{I}$ with rectangles  $R=\{r_1, \ldots, r_n\}$ the bottom-left algorithm places 
the rectangles in the given order at a lowest free position in the strip, using the left most position in case of ties. 
More formally, the bottom-left algorithm will place rectangle $r_1$ at position $(0,0)$. This is a feasible packing of the first rectangle. Assume that the bottom-left algorithm has obtained a feasible packing of the first $i-1$ rectangles into the strip. Then it chooses a position $(x_i, y_i)$ that results in a feasible packing for the first $i$ 
rectangles such that $(y_i, x_i)$ is lexicographically minimal among all possible choices for the position $(x_i, y_i)$.
The height of the packing computed by the bottom-left algorithm on an instance $\cal I$ is denoted by
$\hbl({\cal I})$. This height may heavily depend on the ordering of the rectangles in the instance.
We are therefore also interested in the best possible height that the bottom-left algorithm can achieve for a given set of rectangles, i.e., the minimum height among all $n!$ orderings of the $n$ rectangles. We define this
value as $\hblbest$. Similarly, we define the maximum height among all $n!$ orderings as $\hblworst$. 

The approximation ratio achieved by the bottom-left algorithm on an instance $\cal I$ is defined as the ratio~$\hbl({\cal I}) / \hopt ({\cal I})$. We are also interested in the best possible and worst possible approximation ratio of the
bottom-left algorithm which are defined as $\hblbest({\cal I}) / \hopt ({\cal I})$ and
as $\hblworst({\cal I}) / \hopt ({\cal I})$.

\section{An improved lower bound for the best bottom-left packing}\label{section:improved_lower_bound}
Even for the best possible ordering of rectangles, the bottom-left algorithm might produce a non-optimal packing.
Baker, Coffman, and Rivest \cite{baker:1980} were the first to show that for all $\varepsilon>0$
the bottom-left algorithm cannot have an approximation ratio better than $12/11 - \varepsilon$.
Later an improvement was given by Brown~\cite{brown:1980}, showing that there exists a set of eight rectangles 
for which the bottom-left algorithm cannot have an approximation ratio below $5/4$.
Up to now, this was the best-known lower bound for the ratio between the height of a best bottom-left packing and the height of an  optimal packing.
Furthermore, the value $12/11-\varepsilon$ from~\cite{baker:1980} was the best-known lower bound for SSPP-instances.

In~\cref{subsec:lower_bound_rectangles} we improve the lower bound on the best possible
approximation ratio for the bottom-left algorithm from $5/4$ to $4/3-\varepsilon$ by constructing an SPP-instance 
using only seven rectangles. In our proof we use similar arguments as Brown~\cite{brown:1980}.
We extend this result in~\cref{subsec:lower_bound_squares} to the special case of square packing.
We present a construction for large SSPP instances that yield a lower bound of $4/3-\varepsilon$.
This significantly improves upon the old lower bound of $12/11-\varepsilon$~\cite{baker:1980}. 
Our construction requires substantially novel arguments.

\subsection{Rectangular case}\label{subsec:lower_bound_rectangles}
The main idea in this section to show a better lower bound for the best bottom-left algorithm is to construct an instance that has an optimum packing with bottom-left structure such that the optimum packing is unique up to symmetries.
After that, the instance is slightly modified, preserving the uniqueness (up to symmetry) of the optimum packing, but loosing the bottom-left structure for the optimum packing.
This will result in the desired lower bound.

    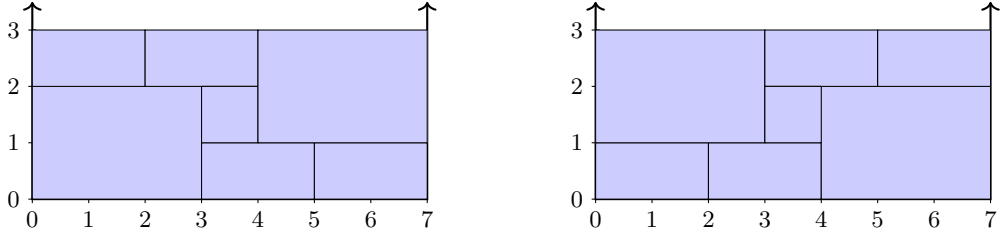
\begin{figure}[t]
        \centering
        \begin{tikzpicture}[scale=0.75]
            \strip73
            \filldraw[myfill] (0,0) rectangle +(3,2);
            \filldraw[myfill] (3,0) rectangle +(2,1);
            \filldraw[myfill] (5,0) rectangle +(2,1);
            \filldraw[myfill] (0,2) rectangle +(2,1);
            \filldraw[myfill] (2,2) rectangle +(2,1);
            \filldraw[myfill] (3,1) rectangle +(1,1);
            \filldraw[myfill] (4,1) rectangle +(3,2);
        \end{tikzpicture}\hspace*{1.5cm}
        \begin{tikzpicture}[scale=0.75]
            \strip73
            \filldraw[myfill] (0,0) rectangle +(2,1);
            \filldraw[myfill] (2,0) rectangle +(2,1);
            \filldraw[myfill] (4,0) rectangle +(3,2);
            \filldraw[myfill] (0,1) rectangle +(3,2);
            \filldraw[myfill] (3,2) rectangle +(2,1);
            \filldraw[myfill] (3,1) rectangle +(1,1);
            \filldraw[myfill] (5,2) rectangle +(2,1);
        \end{tikzpicture}
        \caption{The two optimum solutions for packing the rectangles  $(3,2)$, $(3,2)$, $(2,1)$, $(2,1)$, $(2,1)$, $(2,1)$, $(1,1)$
        into a strip of width~7.}
        \label{fig:opt_packing_4/3}
    \end{figure}
    
\begin{theorem}\label{thm:rec4/3}
    For all $\varepsilon>0$ the approximation ratio of the bottom-left algorithm for the Strip Packing Problem cannot be better than $4/3-\varepsilon$ even if the rectangles are ordered in the best possible way.     
\end{theorem}
\begin{proof}
    Consider an instance with rectangles $(3,2)$, $(3,2)$, $(2,1)$, $(2,1)$, $(2,1)$, $(2,1)$, $(1,1)$ 
    and strip width~$7$.
    The packings  of this instance in~\cref{fig:opt_packing_4/3} with height $3$ are tight, i.e., all space is occupied.
    Hence these packings are optimal. 

    \textbf{Claim}: The packings from~\cref{fig:opt_packing_4/3} are the only optimal packings.
    To prove this claim, consider the three disjoint $1\times 7$ horizontal rows for an arbitrary optimal packing.
    Let the type of a row be a multiset of the sizes of rectangles that the row intersects.
    There are three possible types: (a) $\{3,3,1\}$, (b) $\{3,2,2\}$ and~(c)~$\{2,2,2,1\}$.
    
    Let $a$, $b$, and $c$ denote the number of rows of type~$(a)$, resp.\ type~$(b)$, resp.\ type~$(c)$.
    There is a total of three rows, therefore it holds that $a+b+c = 3$.
    Furthermore, there is exactly one rectangle of width $1$ and this rectangle has height~$1$, hence it holds that $a+c = 1$, thus it follows that $b=2$.
    Now, there are four rectangles of width $2$, these rectangles all have height~$1$, so $2b+3c=4$, this implies that~$c=0$.
    In conclusion, required is that~$a=1$,~$b=2$, and~$c=0$.
    
    Both rectangles of width $3$ have height $2$, hence the row of type $(a)$ must be in the middle, as otherwise there must be another row of type~$(a)$, contradicting that~$a=1$.
    Furthermore, if the square of width $1$ is placed at one of the sides of the strip, then either in the top or bottom row another rectangle of width $1$ is needed.
    As there is no other rectangle of width $1$, the square of width $1$ must be placed in the middle.
    Thus the packings in~\cref{fig:opt_packing_4/3} are the only possible packings with one row of type~$(a)$ and two rows of type~$(b)$.
    This proves the claim.

    Next, we modify the instance slightly so that the two rectangles of width $3$ become a bit thinner, and the square of size $1$ becomes a bit higher.
    More formally, for $\varepsilon>0$ sufficiently small, let $(3-\varepsilon,2)$, $(3-\varepsilon,2)$, $(2,1)$, $(2,1)$, $(2,1),(2,1),(1,1+\varepsilon)$ be the rectangles of the modified instance~$\mathcal{I}_\varepsilon$.
    Let the width of the strip still be~$7$.
    The preceding proof shows that (to within~$\varepsilon)$ the packings of~\cref{fig:opt_packing_4/3} are still optimum and have height $3+\varepsilon$.
    However, for each of the optimal solution, the packing cannot be a bottom-left packing.

    Consider the first packing of~\cref{fig:opt_packing_4/3}, the two top rectangles of size $(2,1)$ 
    have to be packed last in the bottom-left algorithm.
    However, as the $(3,2)$ rectangle shrinks to size $(3-\varepsilon,2)$, there is no space to fit two $(2,1)$ rectangles, unless if the top right $(3-\varepsilon,2)$ is shifted a bit to the right, breaking the bottom-left structure. The packing obtained by the bottom-left algorithm is depicted in ~\cref{fig:problem1_4/3}(a).
    \begin{figure}[ht]
        \centering
        \begin{tikzpicture}[scale=0.61]
            \strip74
            \filldraw[myfill] (0,0) rectangle +(2.8,2);
            \filldraw[myfill] (2.8,0) rectangle +(2,1);
            \filldraw[myfill] (4.8,0) rectangle +(2,1);
            \filldraw[myfill] (0,2) rectangle +(2,1);
            \filldraw[myfill] (0,3) rectangle +(2,1);
            \filldraw[myfill] (2.8,1) rectangle +(1,1.2);
            \filldraw[myfill] (3.8,1) rectangle +(2.8,2);
            \draw (3.5, -1.5) node {(a)};
        \end{tikzpicture}\hspace*{8mm}
             \begin{tikzpicture}[scale=0.61]
            \strip74
            \filldraw[myfill] (0,0) rectangle +(2,1);
            \filldraw[myfill] (2,0) rectangle +(2,1);
            \filldraw[myfill] (4,0) rectangle +(2.8,2);
            \filldraw[myfill] (0,1) rectangle +(2.8,2);
            \filldraw[myfill] (3.8,2) rectangle +(2,1);
            \filldraw[myfill] (2.8,1) rectangle +(1,1.2);
            \filldraw[myfill] (0,3) rectangle +(2,1);
            \draw (3.5, -1.5) node {(b)};
        \end{tikzpicture}\hspace*{8mm}
        \begin{tikzpicture}[scale=0.61]
            \strip74
            \filldraw[myfill] (0,0) rectangle +(2,1);        
            \filldraw[myfill] (2,0) rectangle +(2,1);
            \filldraw[myfill] (4,0) rectangle +(2.8,2);
            \filldraw[myfill] (0,1) rectangle +(2.8,2);
            \filldraw[myfill] (2.8,2) rectangle +(2,1);
            \filldraw[myfill] (4.8,2) rectangle +(1,1.2);
            \filldraw[myfill] (0,3) rectangle +(2,1);
            \draw (3.5, -1.5) node {(c)};
        \end{tikzpicture}\vspace*{-5mm}
        \caption{(a) The bottom-left packing of~$\mathcal{I}_{0.2}$ resulting from the optimum solution shown in the left of~\cref{fig:opt_packing_4/3}. 
        (b)~The bottom-left packing if the rectangle of size $(1,1+\varepsilon)$ is placed before the two top rectangles of size~$(2,1)$ resulting from the optimum solution shown in the right of~\cref{fig:opt_packing_4/3}. 
        (c)~The best possible bottom-left packing if at least one of the two top rectangles of size~$(2,1)$ is placed before the rectangle of size~$(1,1+\varepsilon)$.}
        \label{fig:problem1_4/3}
    \end{figure}
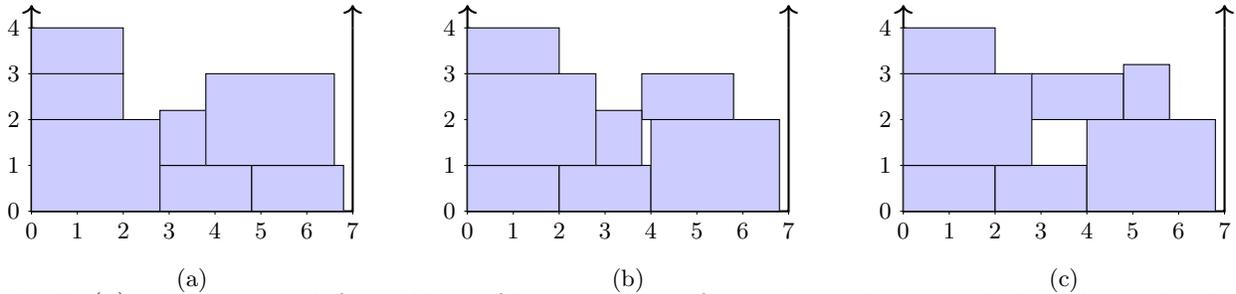
    Next, consider the second packing of~\cref{fig:opt_packing_4/3}.
    If the rectangle of size $(1,1+\varepsilon)$ is packed before the two top right rectangles of size~$(2,1)$, then the bottom-left algorithm returns the packing shown in~\cref{fig:problem1_4/3}(b).
    Else if one of the rectangles of size~$(2,1)$ is placed before the rectangle of size $(1,1+\varepsilon)$, then the best packing that can be obtained by the bottom-left algorithm is shown in~\cref{fig:problem1_4/3}(c).
    
    In conclusion, the best bottom-left packing has height $4$ while an optimum packing has height $3+\varepsilon$. 
    For~$\varepsilon' := 4\varepsilon/(9+3\varepsilon)$ we get a lower bound of $\frac43-\varepsilon'$ on the approximation ratio of the bottom-left algorithm.
    \qed
\end{proof}

The instance from~\cref{thm:rec4/3} can be scaled such that all rectangles have integer side lengths.
Let $h\in \mathbb{N}$ and consider the instance with two rectangles of size $(4,2h)$, four rectangles of size $(3,h)$, one rectangle of size $(1,h+1)$ and let the strip width be $10$.
Using the same arguments as before, the best bottom-left packing has height $4h$ while an optimum packing has height $3h+1$, resulting in a lower bound of~$4h/(3h+1)$.

\subsection{Square case}\label{subsec:lower_bound_squares}
There is an easy modification of the instance from~\cref{thm:rec4/3} such that all rectangles are squares.
In~\cref{cor:6/5} it is shown that this results in a $6/5-\varepsilon$ lower bound for the bottom-left algorithm for SSPP-instances.
After that,~\cref{cor:6/5} is generalized to get a $4/3-\varepsilon$ lower bound for the bottom-left algorithm for SSPP-instances.
\begin{corollary}
\label{cor:6/5}
    For all $\varepsilon>0$ the approximation ratio of the bottom-left algorithm for the Square Strip Packing Problem cannot be better than $6/5-\varepsilon$ even if the squares are ordered in the best possible way.     
\end{corollary}
\begin{proof}
    Consider the collection of squares of sizes $3,3,2,2,2,2,1$.
    Let the width of the strip be~$7$.
    Note that the widths of these squares are the same as the widths of the rectangles from the instance in the proof of~\cref{thm:rec4/3}.
    Using similar reasoning as in the proof of~\cref{thm:rec4/3}, the two packings of~\cref{fig:opt_packing_6/5}(a) 
    and~\cref{fig:opt_packing_6/5}(b) are the only optimal packings.
    \begin{figure}[ht]
        \centering
        \begin{tikzpicture}[scale=0.65]
         \strip75
         \filldraw[myfill] (0,0) rectangle +(3,3);
         \filldraw[myfill] (3,0) rectangle +(2,2);
         \filldraw[myfill] (5,0) rectangle +(2,2);
         \filldraw[myfill] (0,3) rectangle +(2,2);
         \filldraw[myfill] (2,3) rectangle +(2,2);
         \filldraw[myfill] (3,2) rectangle +(1,1);
         \filldraw[myfill] (4,2) rectangle +(3,3);
         \draw (3.5, -1.5) node {(a)};
        \end{tikzpicture}\hspace*{4mm}
        \begin{tikzpicture}[scale=0.65]
            \strip75
            \filldraw[myfill] (0,0) rectangle +(2,2);
            \filldraw[myfill] (2,0) rectangle +(2,2);
            \filldraw[myfill] (4,0) rectangle +(3,3);
            \filldraw[myfill] (0,2) rectangle +(3,3);
            \filldraw[myfill] (3,3) rectangle +(2,2);
            \filldraw[myfill] (3,2) rectangle +(1,1);
            \filldraw[myfill] (5,3) rectangle +(2,2);
        \draw (3.5, -1.5) node {(b)};
        \end{tikzpicture}\hspace*{4mm}
        \begin{tikzpicture}[scale=0.65]
            \strip76
            \filldraw[myfill] (0,0) rectangle +(2.8,2.8);
            \filldraw[myfill] (2.8,0) rectangle +(2,2);
            \filldraw[myfill] (4.8,0) rectangle +(2,2);
            \filldraw[myfill] (0,2.8) rectangle +(2.8,2.8);
            \filldraw[myfill] (2.8,2) rectangle +(2,2);
            \filldraw[myfill] (2.8,4) rectangle +(1.1,1.1);
            \filldraw[myfill] (4.8,2) rectangle +(2,2);
        \draw (3.5, -1.5) node {(c)};
        \end{tikzpicture}\vspace*{-4mm}
        \caption{(a) and (b) show the two optimal packings for the squares of sizes 3, 3, 2, 2, 2, 2, 1 in a strip of width~7. (c) shows the best bottom-left packing for the modified square packing instance ${\cal I}_{0.1}$.}
        \label{fig:opt_packing_6/5}
    \end{figure}
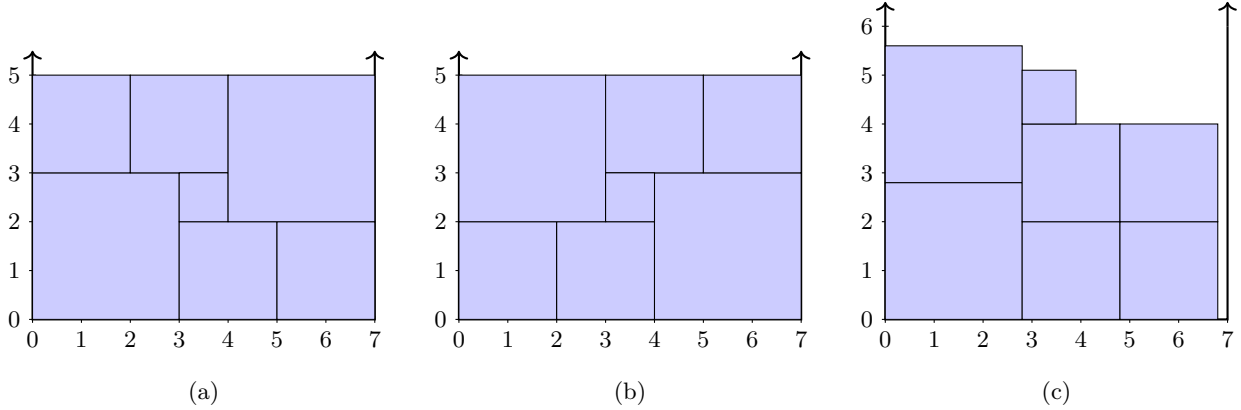
    
    Let $\varepsilon>0$, and consider the modified instance ${\cal I}_\varepsilon$ where the squares of size $3$ get size $3-2\varepsilon$, and the square of size $1$ gets size $1+\varepsilon$.
    Like in~\cref{thm:rec4/3}, the packings from~\cref{fig:opt_packing_6/5}(a) 
    and~\cref{fig:opt_packing_6/5}(b) are still optimum, but do not have the bottom-left structure.

    We may assume $\frac15\ge \varepsilon$. If the squares of size~$3-2\varepsilon$ are on top of each other, then the packing has height at least~$6-4\varepsilon$, and as~\cref{fig:opt_packing_6/5}(c) shows there is a bottom-left packing with height~$6-4\varepsilon$.
    If the squares of size~$3-2\varepsilon$ are not on top of each other, then a bottom-left packing has height at least~$6-\varepsilon$.
    Namely, suppose the height of the packing is less than~$6-\varepsilon$, then each column containing a square of size $3-2\varepsilon$ contains at most one other square of size~$1+\varepsilon$ or size~$2$.
    Furthermore, the squares of size~$3-2\varepsilon$ are either adjacent or non-adjacent.
    If these squares are adjacent, then there must be a column with a square of size~$3-2\varepsilon$ and two squares of size $2$, hence the height of the packing is not less than~$6-\varepsilon$.
    Otherwise, if the squares of size~$3-2\varepsilon$ are not adjacent, then the packing must be as
    in~\cref{fig:opt_packing_6/5}(a) or~\cref{fig:opt_packing_6/5}(b), however, these are not bottom-left packings.
    Thus if the two squares of size $3-2\varepsilon$ are not on top of each other, then the height of a bottom-left packing is at least~$6-\varepsilon$.
    Note that this corresponds to a packing of~\cref{fig:opt_packing_6/5}(a) or~\cref{fig:opt_packing_6/5}(b) where the square of size~$1+\varepsilon$ is placed last on top of the packing.
    Thus the best bottom-left height is $6-4\varepsilon$ while an optimum solution has height $5+\varepsilon$.
    Therefore the ratio equals $(6-4\varepsilon)/(5+\varepsilon)$. Setting $\varepsilon' := 26\varepsilon/(25-5\varepsilon)$ results in the ratio $6/5-\varepsilon'$.
    \qed
\end{proof}

Next~\cref{thm:best_BL_square_lower_bound} will improve the $6/5-\varepsilon$ lower bound by making the instance larger in width, height and number of squares.
After constructing the larger instance, the proof continues similar to the proof of~\cref{thm:rec4/3}.
The resulting lower bound is equal to $4/3-\varepsilon$.
\thmlowerboundsquares*
\begin{proof}
    Let $h\geq 2$ be an integer.
    Consider the instance with one square of size~$h$, $4h$ squares of size~$h+1$ and~$2h$ squares of size~$2h+1$.
    Let the width of the strip be $W = 4h^2 + 3h$.
    The packing in~\cref{fig:sq_opt_4/3} uses $2h$ squares of size $h+1$ on the bottom left, $h$ squares of size~$2h+1$ on the top left, and conversely $2h$ squares of size $h+1$ on the top right, and $h$ squares of size $2h+1$ on the bottom right.
    Thus, this leaves a gap in the middle for the square of size $h$.
    The packing is tight and hence the optimal packing has height $3h+2$.
    
    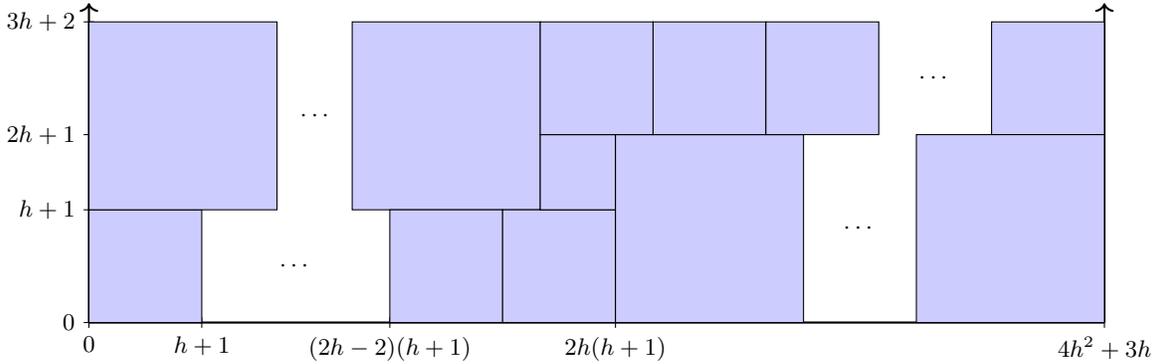
\begin{figure}[ht]
        \centering
        \begin{tikzpicture}[scale=0.5]
            \draw[thick] (0,0) -- (27,0) node[anchor=north west] {};
            \draw[thick,->] (0,0) -- (0,8.5) node[anchor=south east] {};
            \draw[thick,->] (27,0) -- (27,8.5) node[anchor=south east] {};

            \draw (4pt,0) -- (-4pt,0) node[anchor=east] {$0$};
            \draw (4pt,3) -- (-4pt,3) node[anchor=east] {$h+1$};
            \draw (4pt,5) -- (-4pt,5) node[anchor=east] {$2h+1$};
            \draw (4pt,8) -- (-4pt,8) node[anchor=east] {$3h+2$};

            \draw (0,4pt) -- (0,-4pt) node[anchor=north] {$0$};
            \draw (3,4pt) -- (3,-4pt) node[anchor=north] {$h+1$};
            \draw (8,4pt) -- (8,-4pt) node[anchor=north] {$(2h-2)(h+1)$};
            \draw (14,4pt) -- (14,-4pt) node[anchor=north] {$2h(h+1)$};
            \draw (27,4pt) -- (27,-4pt) node[anchor=north] {$4h^2 +3h$};

            \draw (5.5,1.5) node[] {$\cdots$};
            \draw (6.05,5.5) node[] {$\cdots$};
            \draw (20.5,2.5) node[] {$\cdots$};
            \draw (22.5,6.5) node[] {$\cdots$};

            \filldraw[myfill]  (0,0) rectangle +(3,3);
            \filldraw[myfill]  (0,3) rectangle +(5,5);
            \filldraw[myfill]  (7,3) rectangle +(5,5);
            \filldraw[myfill]  (8,0) rectangle +(3,3);
            \filldraw[myfill] (11,0) rectangle +(3,3);
            \filldraw[myfill] (12,3) rectangle +(2,2);
            \filldraw[myfill] (14,0) rectangle +(5,5);
            \filldraw[myfill] (12,5) rectangle +(3,3);
            \filldraw[myfill] (15,5) rectangle +(3,3);
            \filldraw[myfill] (18,5) rectangle +(3,3);
            \filldraw[myfill] (24,5) rectangle +(3,3);
            \filldraw[myfill] (22,0) rectangle +(5,5);
        \end{tikzpicture}\vspace*{-2mm}
        \caption{The unique optimal square packing up to symmetry.}
        \label{fig:sq_opt_4/3}
    \end{figure}

    Consider the $4h^2+3h$ disjoint $1 \times (3h+2)$ vertical columns.
    There are two possible types $(a)$ $\{h+1,h+1,h\}$ and $(b)$ $\{2h+1,h+1\}$.
    Let $a$ respectively $b$ denote their number.
    In total there are $4h^2+3h$ columns, hence it holds that $a+b = 4h^2+3h$.
    Furthermore, there is only one square of size $h$, thus $a=h$.
    This implies that~$b = 4h^2+2h$.

    From $a = h$ it follows that either the square of size~$h$ is above (symmetrically below) two rows of squares of size $h+1$ or it is between squares of size $h+1$.
    The first case is not possible, as this creates a gap of height~$h$ which cannot be filled by another square, this is demonstrated by the red area in~\cref{fig:abc_4/3_sq_problem}(a).
    In the second case, either the two squares of size~$h+1$ above and below the square of size~$h$ go over the same side of the square of size $h$ as in~\cref{fig:abc_4/3_sq_problem}(b), then no other square can fill the red space, hence this is not possible.
    Thus the square above and below the square of size~$h$ must go in different directions over the left respectively right boundary of the square of size~$h$, as depicted in~\cref{fig:abc_4/3_sq_problem}(c).
    As $b = 4h^2+2h$, it follows that each other column with a square of size $h+1$ also contains a square of size $2h+1$.
    Thus the structure in~\cref{fig:abc_4/3_sq_problem}(d) must be part of the optimal solution.

    \begin{figure}[ht]
    \vspace*{-4mm}
        \centering
        \begin{tikzpicture}[scale=0.48]
            \draw[thick,->] (0,0) -- (29.5,0) node[anchor=north west] {};
            \draw[thick,->] (0,0) -- (0,8.5) node[anchor=south east] {};
            \draw (4pt,0) -- (-4pt,0) node[anchor=east] {\footnotesize$0$};
            \draw (4pt,3) -- (-4pt,3) node[anchor=east] {\footnotesize$h+1$};
            \draw (4pt,5) -- (-4pt,5) node[anchor=east] {\footnotesize$2h+1$};
            \draw (4pt,6) -- (-4pt,6) node[anchor=east] {\footnotesize$2h+2$};
            \draw (4pt,8) -- (-4pt,8) node[anchor=east] {\footnotesize$3h+2$};

            \filldraw[myfill] (1,0) rectangle +(3,3);
            \filldraw[myfill] (1,3) rectangle +(3,3);
            \filldraw[myfill] (1,6) rectangle +(2,2);
            \filldraw[myredfill] (3,6) rectangle +(1,2);   
            \draw (2.5, -1.5) node {(a)};

            \filldraw[myfill] (6,0) rectangle +(3,3);
            \filldraw[myfill] (6,3) rectangle +(2,2);
            \filldraw[myfill] (6,5) rectangle +(3,3);
            \filldraw[myredfill] (8,3) rectangle +(1,2);    
            \draw (7.5, -1.5) node {(b)};

            \filldraw[myfill] (11,0) rectangle +(3,3);
            \filldraw[myfill] (12,3) rectangle +(2,2);
            \filldraw[myfill] (12,5) rectangle +(3,3);
            \draw (12.5, -1.5) node {(c)};

            \filldraw[myfill] (17,3) rectangle +(5,5);
            \filldraw[myfill] (21,0) rectangle +(3,3);
            \filldraw[myfill] (22,3) rectangle +(2,2);
            \filldraw[myfill] (24,0) rectangle +(5,5);
            \filldraw[myfill] (22,5) rectangle +(3,3);
            \draw (23, -1.5) node {(d)};
        \end{tikzpicture}\vspace*{-3mm}
        \caption{(a) shows that if the square of size $h$ is above two rows of squares of size~$h+1$, then there is no other square that can fill the red area.
        (b) shows that if the square of size~$h$ is between squares of size $h+1$, then there is no other square that can fill the red area.
        The only possibility where the square of size $h$ is between squares of size~$h+1$ is (up to symmetry) 
        shown in (c). Up to symmetry the structure shown in (d) must be part of the optimal packing.}
        \label{fig:abc_4/3_sq_problem}
    \end{figure}
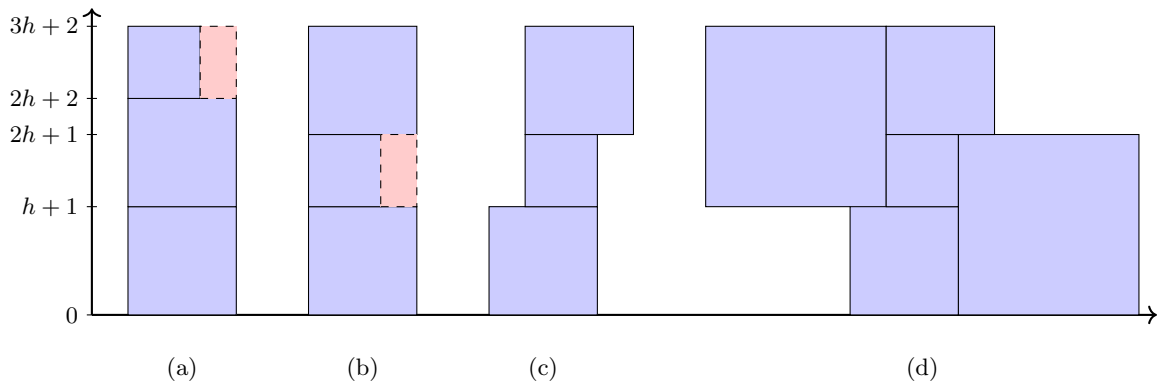

    Next, building on the structure in~\cref{fig:abc_4/3_sq_problem}(d), $b-2$ more columns of type~$(b)$ are required.
    As the optimal packing is tight, it follows that next to the square of size $h+1$ in the bottom left, there must be other squares of size $h+1$, and similar next to the square of size~$2h+1$ in the top left, there must be other squares of size~$2h+1$.
    More precisely, there must be at least $2h-1$ squares of size $h+1$ next to the square of size $h+1$ and at least $h-1$ squares of size $2h+1$ next to the square of size $2h+1$, because those are the smallest numbers such that the left side of the left most square of size $h+1$ is at the same place as the left side of the left most square of size $2h+1$.
    In other words, the width of placing $2h$ squares of size $h+1$ next to each other equals $2h(h+1)$, which is the same as the width of placing $h$ squares of size $2h+1$ next to each other together with the width of the square of size $h$, that is, $h(2h+1) + h$.
    The same holds for the right side of the structure in~\cref{fig:abc_4/3_sq_problem}(d).
    As there are only~$4h$ squares of size $h+1$ and $2h$ squares of size $2h+1$, it follows that the packing in~\cref{fig:sq_opt_4/3} (up to symmetry) is the unique optimal packing.

    Let $\varepsilon>0$ be sufficiently small and consider the slight modification where the squares of size~$2h+1$ have size~$2h+1-\varepsilon$ and the square of size $h$ has size $h+\varepsilon$.
    The packing of~\cref{fig:sq_opt_4/3} has height $3h + 2 -\varepsilon$ and still is the unique optimal packing (up to symmetry), but it does not have the bottom-left structure.
    Now, similar to~\cref{cor:6/5}, the packing in~\cref{fig:BL_sq_4/3} is an optimal bottom-left packing and has height $4h+2-2\varepsilon$.
    Because if the squares of size $2h+1-\varepsilon$ are not on top of each other, then the height of a bottom-left packing is at least $4h+2$, because there is a column containing the squares of size~$2h+1-\varepsilon$, $h+1$ and $h+\varepsilon$.
    
    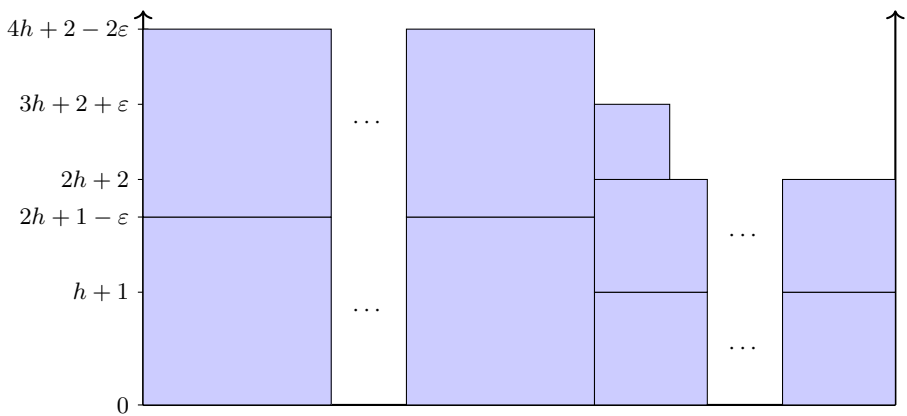
\begin{figure}[ht]
        \centering
        \begin{tikzpicture}[scale=0.5]
            \draw[thick] (0,0) -- (20,0) node[anchor=north west] {};
            \draw[thick,->] (0,0) -- (0,10.5) node[anchor=south east] {};
            \draw[thick,->] (20,0) -- (20,10.5) node[anchor=south east] {};
            \draw (4pt,0) -- (-4pt,0) node[anchor=east] {\footnotesize$0$};
            \draw (4pt,3) -- (-4pt,3) node[anchor=east] {\footnotesize$h+1$};
            \draw (4pt,5) -- (-4pt,5) node[anchor=east] {\footnotesize$2h+1-\varepsilon$};
            \draw (4pt,6) -- (-4pt,6) node[anchor=east] {\footnotesize$2h+2$};
            \draw (4pt,8) -- (-4pt,8) node[anchor=east] {\footnotesize$3h+2+\varepsilon$};
            \draw (4pt,10) -- (-4pt,10) node[anchor=east] {\footnotesize$4h+2-2\varepsilon$};

            \draw (6,2.5) node[] {$\cdots$};
            \draw (6,7.5) node[] {$\cdots$};

            \draw (16,1.5) node[] {$\cdots$};
            \draw (16,4.5) node[] {$\cdots$};

            \filldraw[myfill] (0,0) rectangle +(5,5);
            \filldraw[myfill] (0,5) rectangle +(5,5);
            \filldraw[myfill] (7,0) rectangle +(5,5);
            \filldraw[myfill] (7,5) rectangle +(5,5);
            \filldraw[myfill] (12,0) rectangle +(3,3);
            \filldraw[myfill] (12,3) rectangle +(3,3);
            \filldraw[myfill] (17,0) rectangle +(3,3);
            \filldraw[myfill] (17,3) rectangle +(3,3);
            \filldraw[myfill] (12,6) rectangle +(2,2);
           
        \end{tikzpicture}
        \vspace*{-3mm}
        \caption{An optimum bottom-left packing of the modified instance.}
        \label{fig:BL_sq_4/3}
    \end{figure}

    Thus for sufficiently small $\varepsilon>0$, and every $h\geq 2$, there exists an instance~$\mathcal{I}_{\varepsilon,h}$ such that
    \begin{align*}
        \frac{h_{\text{BL}}^{\text{best}}(\mathcal{I}_{\varepsilon,h})}{h_{\text{OPT}}(\mathcal{I}_{\varepsilon,h})} ~=~ \frac{4h-2\varepsilon+2}{3h+2-\varepsilon}.
    \end{align*}
    Setting $\epsilon' := (2+2\varepsilon)/(9h+6-3\varepsilon)$ we get a lower bound of $4/3-\varepsilon'$
    on the approximation ratio of the bottom-left algorithm in case of squares. As $\varepsilon'$ gets arbitrarily small for sufficiently large $h$ this proves the result. 
    \qed
\end{proof}

Consequently, even when $90$-degree rotation of rectangles is allowed, there exists an instance such that the ratio between the height of a bottom-left packing using the best ordering and the height of an optimal solution is at least $4/3-\varepsilon$.
Namely, the instance from~\cref{thm:best_BL_square_lower_bound} is a set of rectangles that is invariant under rotation, because each rectangle is a square.

\section{Bounds for the worst-order bottom-left packing}\label{section:worst_case_BL}
If the rectangles are badly ordered, the ratio between the height of a bottom-left packing and the height of an optimal packing can be arbitrarily large. This might even happen if the rectangles
are ordered by increasing width~\cite{baker:1980}.
On the contrary, for SSPP-instances we can be show that the approximation ratio of the bottom-left algorithm remains bounded regardless of the ordering of the squares and also improve the so far best-known lower bound of $2-\varepsilon$~\cite{baker:1980} for this case:

\thmworstcaseordering*

\begin{proof}
Our proof for the constant upper bound is very similar to a proof presented in~\cite{fekete2014online}
to show that the online \textit{BottomLeft} algorithm has approximation ratio $3.5$. 
The main idea is to prove that the holes in a bottom-left packing can be covered by surrounding squares.
Contrary to~\cite{fekete2014online} we do not aim to get the best possible constant.
This makes our proof shorter than the one presented in~\cite{fekete2014online} at the cost of getting a worse
constant ($16$ instead of $3.5$). We present the details of our proof for the upper bound 
in the appendix. 

To prove the lower bound of  $10/3-\varepsilon$ we start 
with the so-called checkerboard construction from~\cite{baker:1980}.

\begin{construction}[$m$-checkerboard]\label{construction:checkerboard}
    Let $m\geq 2$ be even and define $\varepsilon = \frac{2}{m^3(m^2+1)}$.
    Consider the instance consisting of squares of size~$2- i \varepsilon$ for~$i=1,\dots,m^2$ together with~$m^3 + \frac{(m-1)m}{2}$ unit squares.
    Let the strip width be
    \begin{align*}
        W = \sum_{i=1}^{m^2} (2-i\varepsilon) ~=~  2m^2 - \varepsilon \frac{m^2(m^2+1)}{2} ~=~  2m^2 - \frac{1}{m}.
    \end{align*}

    The bottom-left packing of the squares ordered by decreasing size is shown in~\cref{fig:checkerboard} (blue squares).
    By definition, all large squares fit precisely next to each other on the strip bottom.
    Furthermore, these squares are ordered by decreasing size, hence the unit squares on top of these large squares are placed from right to left.
    Since two unit squares are wider than any of the large squares, it follows that only one unit square is placed on top of each large square.
    This type of placement repeats in the second row because the gaps between the unit squares in the first row all have width less than~$1$.
    In general, the $i$-th row of unit squares alternates gaps and squares except for the initial and final squares of the row, these form almost solid triangles.
    More precisely, the $i$-th row contains $m^2+i-1$ unit squares, and there are precisely $m$ such rows, because
    \begin{align*}
        \sum_{i=1}^m (m^2 + i - 1) ~=~ m^3 + \frac{(m-1)m}{2}.
    \end{align*}
    Last of all, each row fits into the strip. 
    The $i$-th row consists of $m^2+(i-1)$ unit squares and $m^2-i$ gaps between these squares of size less than~$1$, hence each row has width less than $2m^2-1 < W$.
    From this we conclude that the bottom-left packing as depicted in~\cref{fig:checkerboard} (blue squares) is correct.
    Moreover, the height of this bottom-left packing equals $m + 2 - \varepsilon$.

    \begin{figure}[ht]
        \vspace*{-5mm}
        \centering
        \begin{tikzpicture}[scale=0.46]
            \def\eps{0.001838}
            \def\m{4}
            \def\W{2 * \m * \m - 0.25}
            \draw[thick,->] (0,0) -- (0,7.6) node[anchor=south east] {};
            \draw[thick] (0,0) -- (\W,0) node[anchor=north west] {};
            \draw[thick,->] (\W, 0) -- (\W, 7.6) node[anchor=south east] {};
            
            \foreach \x in {1,...,16}
                \filldraw[myfill] (2*\x - 2 - 0.5*\x*\x*\eps + 0.5*\x*\eps, 0) rectangle +(2 - \x*\eps, 2 - \x*\eps);   
                
            \foreach \x in {1,...,16}
                \filldraw[myfill] (2*\x - 2 - 0.5*\x*\x*\eps + 0.5*\x*\eps, 2 - \x*\eps) rectangle +(1, 1);
                
            \foreach \x in {1,...,15}
                \filldraw[myfill] (2*\x - 1 - 0.5*\x*\x*\eps + 0.5*\x*\eps, 3 - \x*\eps - \eps) rectangle +(1, 1);
                
            \filldraw[myfill] (2*15 - 0.5*15*15*\eps + 0.5*15*\eps, 3 - 15*\eps - \eps) rectangle +(1, 1);
            \foreach \y in {0,...,2}
                \filldraw[myfill] (0, \y + 3 - \eps) rectangle +(1, 1);

            \foreach \x in {1,...,14}
                \filldraw[myfill] (2*\x - 0 - 0.5*\x*\x*\eps + 0.5*\x*\eps, 4 - \x*\eps - 2*\eps) rectangle +(1, 1);

            \foreach \x in {1,2}    
                \filldraw[myfill] (\x + 2*14 - 0.5*14*14*\eps + 0.5*14*\eps, 4 - 14*\eps - 2*\eps) rectangle +(1, 1);
            \foreach \y in {0,...,1}
                \filldraw[myfill] (1, \y + 4 - 2*\eps) rectangle +(1, 1);
 
            \foreach \x in {1,...,13}
                \filldraw[myfill] (2*\x + 1 - 0.5*\x*\x*\eps + 0.5*\x*\eps, 5 - \x*\eps - 3*\eps) rectangle +(1, 1);
            
            \foreach \x in {1,2,3}    
              \filldraw[myfill] (\x + 2*13 + 1 - 0.5*13*13*\eps + 0.5*7*\eps, 5 - 13*\eps - 3*\eps) rectangle +(1, 1);
            \foreach \y in {0}
                \filldraw[myfill] (2, \y + 5 - 3*\eps) rectangle +(1, 1);

            \filldraw[mygreenfill] (0, 6 - 1*\eps) rectangle +(1+1*\eps, 1+1*\eps);
            \filldraw[mygreenfill] (2 - 0* \eps, 6 - 3*\eps) rectangle +(1+3*\eps, 1+3*\eps);
            
            \filldraw[mygreenfill] (4 - 0* \eps, 6 - 4*\eps ) rectangle +(1+5*\eps, 1+5*\eps);
            \filldraw[mygreenfill] (6 - 1* \eps, 6 - 5*\eps) rectangle +(1+6*\eps, 1+6*\eps);
            \filldraw[mygreenfill] (8 - 3* \eps, 6 - 6*\eps ) rectangle +(1+7*\eps, 1+7*\eps);
            \filldraw[mygreenfill] (10 - 6* \eps, 6 - 7*\eps) rectangle +(1+8*\eps, 1+8*\eps);
            \filldraw[mygreenfill] (12 - 10* \eps, 6 - 8*\eps ) rectangle +(1+9*\eps, 1+9*\eps);
            \filldraw[mygreenfill] (14 - 15* \eps, 6 - 9*\eps) rectangle +(1+10*\eps, 1+10*\eps);
            \filldraw[mygreenfill] (16 - 21* \eps, 6 - 10*\eps ) rectangle +(1+11*\eps, 1+11*\eps);
            \filldraw[mygreenfill] (18 - 28* \eps, 6 - 11*\eps ) rectangle +(1+12*\eps, 1+12*\eps);
            \filldraw[mygreenfill] (20 - 36* \eps, 6 - 12*\eps ) rectangle +(1+13*\eps, 1+13*\eps);
            \filldraw[mygreenfill] (22 - 45* \eps, 6 - 13*\eps ) rectangle +(1+14*\eps, 1+14*\eps);
            \filldraw[mygreenfill] (24 - 55* \eps, 6 - 14*\eps ) rectangle +(1+15*\eps, 1+15*\eps);

            \filldraw[mygreenfill] (26 - 66* \eps, 6 - 15*\eps ) rectangle +(1+16*\eps, 1+16*\eps);
            \filldraw[mygreenfill] (27 - 50* \eps, 6 - 15*\eps ) rectangle +(1+16*\eps, 1+16*\eps);
            \filldraw[mygreenfill] (28 - 44* \eps, 6 - 15*\eps ) rectangle +(1+16*\eps, 1+16*\eps);
            \filldraw[mygreenfill] (29 - 28* \eps, 6 - 15*\eps ) rectangle +(1+16*\eps, 1+16*\eps);
            \filldraw[mygreenfill] (30 - 12* \eps, 6 - 15*\eps ) rectangle +(1+16*\eps, 1+16*\eps);

            \draw (4pt,6) -- (-4pt,6) node[anchor=east] {\footnotesize$m+2-\varepsilon$};
            \draw (4pt,7) -- (-4pt,7) node[anchor=east] {\footnotesize$m+3$};
        \end{tikzpicture}
        \vspace*{-6mm}
        \caption{In blue the $m$-checkerboard example for $m=4$, together with the reset row in green.}
        \label{fig:checkerboard}
    \end{figure}
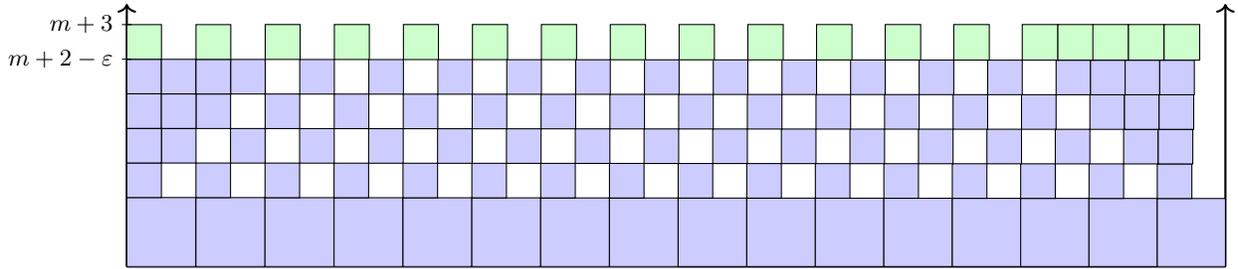

    Next, to construct an upper bound on the height of an optimum packing, observe that $2m^2-1$ unit squares fit next to each other in one row.
    Hence, all unit squares fit into $m/2+1$ rows of height~$1$, because
    $(2m^2-1)(m/2+1)= m^3 + 2m^2 - m/2 - 1 > m^3 + (m-1)m/2$.
    The squares of size $2-i\varepsilon$ fit into one row of height~$2$, thus it holds that $\hopt \leq m/2 + 1 + 2 = m/2+3$.
    Consequently, the bottom-left algorithm for the $m$-checkerboard has approximation ratio at least $\frac{m+2 - \varepsilon}{m/2 + 3}$ which approaches $2$ as $m$ becomes large.
\end{construction}

Adding one large square of size $m/2$ at the end to the $m$-checkerboard example already results in a lower bound approaching~$3$ for the worst-order bottom-left algorithm.
Furthermore, the lower bound can be improved to $10/3-\varepsilon$ when adding a construction with density $\frac{1}{3}$ to the checkerboard and ending with one large square.
This $\frac{1}{3}$-dense construction requires a flat foundation on top of the checkerboard, therefore we add one extra row of squares to the checkerboard such that the height of the top face of these squares is the same and no large gaps exist in this row.

\begin{construction}[Reset row]\label{construction:reset_row}
    Let $m$ be even.
    There are $m^2+m-1$ unit squares in the $m$-th row of the $m$-checkerboard from Construction~\ref{construction:checkerboard}.
    Let $h_i$ be the height of the top face of the $i$-th square in the $m$-th row (from left to right).
    It holds that
    \begin{align*}
        h_i ~=~ \begin{cases}
            m+2-i\varepsilon &\text{if } 1\leq i \leq m^2,\\
            m+2 - m^2\varepsilon &\text{if } m^2 < i < m^2+m,
        \end{cases}
    \end{align*}
    because the $i$-th such unit square leans onto the $i$-th bottom row square of size $2-i\varepsilon$.
    The reset row consists of $m+1$ squares of size $1+m^2\varepsilon$, and a square of size $1+i\varepsilon$ for each $m < i < m^2$ and odd $1\leq i\leq m$.
    Bottom-left place them in decreasing size order on top of the $m$-checkerboard, then the square of size $1+i\varepsilon$ is going to be placed on top of $h_{i}$, hence the height of the top face of these almost unit squares equals
    \begin{align*}
        h_{i} + (1+i\varepsilon) ~=~ (m+2 -i\varepsilon) + (1+i\varepsilon)  ~=~ m+3.
    \end{align*}
    Thus the top face of all squares in the reset row have the same height.
    Moreover, all gaps between squares in this construction have width less than~$1$.
    \cref{fig:checkerboard} depicts the reset row in green.
\end{construction}

    The main idea to obtain the $10/3-\varepsilon$ lower bound is to construct exponentially growing rows above the $m$-checkerboard with gaps of approximately twice the width of the squares in that row. 
    To this end, consider an integer $n\geq 1$ and let $m$ be the largest even number such that $m \leq \frac{4}{3} 2^n$.
    Consider the $m$-checkerboard together with the reset row from Construction~\ref{construction:reset_row}.
    After that, bottom-left pack the pattern consisting of four unit squares followed by one square of size $2+(a_{i}+1) \varepsilon$ for $1\leq i\leq \lfloor \frac{W}{6}\rfloor$ where $a_{i}=1$ if $i$ is odd and $a_{i} = 0$ otherwise.
    These squares fit into one row on top of the reset row and the gaps between these squares is~$4$.
    If the space at the end of this row is at least $4+2\varepsilon$, then add more squares of size $2+\varepsilon$ at the end.
    Next, for~$2\leq j\leq n-1$, bottom-left pack squares of size $2^j+(a_{i}+2^{j-1})\varepsilon$ for $1\leq i \leq \lfloor \frac{W}{2^j+2^{j+1}}\rfloor$.
    Each $j$ constitutes a row because the gaps between the squares in the previous row have width less than
    \begin{align*}
        (2^{j-1} + (1+2^{j-1})\varepsilon) + (2^{j-1} + 2^{j-1}\varepsilon) + (2^j + j\varepsilon) - 2^j\varepsilon ~=~ 2^{j+1} + (j+1)\varepsilon
    \end{align*}
    which can be shown by induction together with the observation that the squares in the $j$-th row are placed on the squares in the $(j-1)$-th row for which $a_i = 0$.
    If the space at the end of a row is larger than $2^{j+1}+2^j\varepsilon$, then add extra squares of size $2^j+2^{j-1}\varepsilon$ to this row. 
    After placing the $n-1$ rows, finish with bottom-left placing one square of size $2^n+2^{n-1}\varepsilon$ on top of the packing.
    This bottom-left packing is depicted in \cref{fig:10over3lowerbound}.
    A lower bound on the height of this bottom-left packing is
    \begin{align*}
        \hblworst ~\geq~ m + \sum_{i=1}^{n-1} 2^i + 2^n ~\geq~ \frac{10}{3} \cdot 2^n,
    \end{align*}
    because the $m$-checkerboard has height at least $\frac{4}{3} 2^n$, the squares in the middle of the structure of Fig.~\ref{fig:10over3lowerbound} have height at least~$\sum_{i=1}^{n-1} 2^i$, and on top of that a square of size at least~$2^n$ is placed.

    As mentioned, the gap between the squares of size $2^j+(a_i+2^{j-1})\varepsilon$ is at most $2^{j+1}+(j+1)\varepsilon$, hence each row is approximately for $\frac{1}{3}$ occupied by squares.
    Furthermore, the checkerboard is approximately half-occupied, and the top row with only a square of size $2^n+2^{n-1}\varepsilon$ is almost entirely empty as $n$ becomes large, because the width of the strip is quadratic in $m$.
    Thus the total amount of occupied area is approximately
    \begin{align*}
        \frac12 \cdot \frac{4}{3} 2^n + \frac{1}{3} \cdot \sum_{i=1}^{n-1} 2^i + 0 \cdot 2^n ~=~ \frac{2}{3} 2^n + \frac13 2^n + \mathcal{O}(1) ~=~ 2^n + \mathcal{O}(1).
    \end{align*}
    As the squares get exponentially small, it is not difficult to construct an optimum packing of height $2^n + \mathcal{O}(1)$.
    Hence there exists $\varepsilon>0$ such that the lower bound for the worst ordering in the bottom-left algorithm is
    \begin{align*}
        \frac{\hblworst}{\hopt} ~\geq~ \frac{\frac{10}{3}\cdot 2^n}{2^n + \mathcal{O}(1)} ~=~ \frac{10}{3} - \varepsilon.
    \end{align*}
\qed

\begin{figure}[ht]
        \centering
        \vspace*{-5mm}
        \begin{tikzpicture}[scale=0.38]
            \def\eps{0.05}
            \def\W{37.8}
            \def\h{17}
            \draw[thick,->] (0,0) -- (0,\h) node[anchor=south east] {};
            \draw[thick] (0,0) -- (\W,0) node[anchor=north west] {};

            \filldraw[myfill] (0, 0) rectangle +(1,1);   
            \filldraw[myfill] (1, 0) rectangle +(1,1);   
            \filldraw[myfill] (2, 0) rectangle +(1,1);   
            \filldraw[myfill] (3, 0) rectangle +(1,1);  
            \filldraw[myfill] (4, 0) rectangle +(2 + 2*\eps, 2 + 2*\eps);   
            \filldraw[myfill] (6 + 2*\eps, 0) rectangle +(1,1);  
            \filldraw[myfill] (7 + 2*\eps, 0) rectangle +(1,1);  
            \filldraw[myfill] (8 + 2*\eps, 0) rectangle +(1,1);  
            \filldraw[myfill] (9 + 2*\eps, 0) rectangle +(1,1);  
            \filldraw[myfill] (10 + 2*\eps, 0) rectangle +(2 + \eps, 2 + \eps);   
            \filldraw[myfill] (12 + 3*\eps, 0) rectangle +(1,1);   
            \filldraw[myfill] (13 + 3*\eps, 0) rectangle +(1,1);   
            \filldraw[myfill] (14 + 3*\eps, 0) rectangle +(1,1);   
            \filldraw[myfill] (15 + 3*\eps, 0) rectangle +(1,1);   
            \filldraw[myfill] (16 + 3*\eps, 0) rectangle +(2 + 2*\eps, 2 + 2*\eps);   
            \filldraw[myfill] (18 + 5*\eps, 0) rectangle +(1,1);   
            \filldraw[myfill] (19 + 5*\eps, 0) rectangle +(1,1);   
            \filldraw[myfill] (20 + 5*\eps, 0) rectangle +(1,1);   
            \filldraw[myfill] (21 + 5*\eps, 0) rectangle +(1,1);   
            \filldraw[myfill] (22 + 5*\eps, 0) rectangle +(2 + \eps, 2 + \eps);   
            \filldraw[myfill] (24 + 6*\eps, 0) rectangle +(1,1);   
            \filldraw[myfill] (25 + 6*\eps, 0) rectangle +(1,1);   
            \filldraw[myfill] (26 + 6*\eps, 0) rectangle +(1,1);   
            \filldraw[myfill] (27 + 6*\eps, 0) rectangle +(1,1);   
            \filldraw[myfill] (28 + 6*\eps, 0) rectangle +(2 + 2*\eps, 2 + 2*\eps);   
            \filldraw[myfill] (30 + 8*\eps, 0) rectangle +(1,1);   
            \filldraw[myfill] (31 + 8*\eps, 0) rectangle +(1,1);   
            \filldraw[myfill] (32 + 8*\eps, 0) rectangle +(1,1);   
            \filldraw[myfill] (33 + 8*\eps, 0) rectangle +(1,1);   
            \filldraw[myfill] (34 + 8*\eps, 0) rectangle +(2 + \eps, 2 + \eps);   
            \filldraw[myfill] (36 + 9*\eps, 0) rectangle +(1,1);

            \filldraw[myfill] (6 + 2*\eps, 2 + \eps) rectangle +(4 + 3*\eps,4 + 3*\eps);
            \filldraw[myfill] (18 + 5*\eps, 2 + \eps) rectangle +(4 + 2*\eps, 4 + 2*\eps);
            \filldraw[myfill] (30 + 8*\eps, 2 + \eps) rectangle +(4 + 3*\eps, 4 + 3*\eps);

            \filldraw[myfill] (10+5*\eps, 6 + 3*\eps) rectangle +(8 + 5*\eps, 8 + 5*\eps);
            \filldraw[myfill_noborder] (34+11*\eps, 6 + 3*\eps) rectangle +(2,8+ 4*\eps);
            \draw[black] (34+11*\eps, 6 + 3*\eps) -- (34+11*\eps+2, 6 + 3*\eps);
            \draw[black] (34+11*\eps, 6 + 3*\eps) -- (34+11*\eps, 6 +3*\eps+8+4*\eps);
            \draw[black] (34+11*\eps, 6 + 3*\eps+8+4*\eps) -- (34+11*\eps+2, 6 + 3*\eps+8+4*\eps);

            \draw (34+11*\eps+2.8, 4+6+4*\eps)  node[] {$\cdots$};

            \filldraw[myfill_noborder] (18+10*\eps, 14 +7*\eps) rectangle +(16+9*\eps,2);
            \draw[black] (18+10*\eps,14 + 7*\eps) -- (18+10*\eps,14 + 7*\eps+2);
            \draw[black] (18+10*\eps,14 + 7*\eps) -- (18+10*\eps+16+9*\eps,14 + 7*\eps);
            \draw[black] (18+10*\eps+16+9*\eps,14 + 7*\eps) -- (18+10*\eps+16+9*\eps,14 + 7*\eps+2);
            

            \draw (16*0.5+18+10*\eps, 14 +7*\eps+3) node[] {$\vdots$};
        \end{tikzpicture}
        \vspace*{-2mm}
        \caption{Exponentially growing squares of size ~$2^j+(a_i+2^{j-1})\varepsilon$ and gaps of size at most $2^{j+1}+(j+1)\varepsilon$.}
        \label{fig:10over3lowerbound}
    \end{figure}
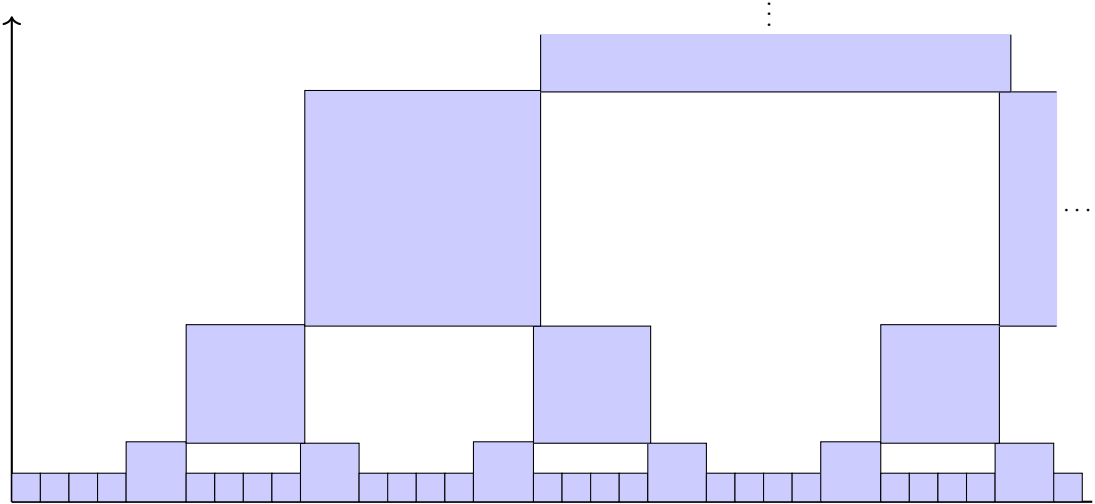
\end{proof}

The lower bound also holds in an online context, where the squares are placed in the order given, without having all squares available from the start.
In this online setting, a $3.5$-upper bound on the competitive ratio of the bottom-left algorithm is known that additionally satisfies the Tetris and gravity constraints~\cite{fekete2014online}.
The Tetris constraint requires that when a square is placed, there is a path from the top of the strip to the final position along which the square can be moved without intersecting the already placed squares.
Furthermore, the gravity constraint restricts this path from going up.
The best-known lower bound for the bottom-left algorithm with Tetris and gravity constraints was $3/2$~\cite{fekete2014online}. 
Our $10/3-\varepsilon$ lower bound also applies to the Tetris-gravity bottom-left algorithm, closing the gap to the $3.5$-upper bound significantly.

\corlowerboundfekete*

\section{Bottom-left $k$-local search}\label{section:local_search}
The bottom-left $k$-local search algorithm computes the height of the bottom-left packing of an instance and tries to improve the height by permuting at most~$k$ rectangles in the ordering and comparing if the new bottom-left packing has a smaller height. 
The algorithm continues until a local optimum is reached, i.e., no more improvement steps exist.
We denote by $\hkbl(\mathcal{I})$ the height of a solution returned by this algorithm for an instance $\mathcal{I}$.
It turns out that there are (Square) Strip Packing instances with approximation ratio at least~$2$ when starting with a bad ordering.

\begin{theorem}
    For every $k\in\mathbb{N}$, there exists a (S)SPP-instance~$\mathcal{I}$ such that $\hkbl(\mathcal{I}) = 2\cdot  h_{\text{OPT}}(\mathcal{I})$.
\end{theorem}
\begin{proof}
    The idea is to have one row of alternating small and big squares followed by one big square, such that for each permutation of~$k$ squares, always one big square is on top, while the optimal packing has all the big squares in the same row.
    More precisely, consider the instance consisting of $2k+4$ unit squares and~$2k+5$ squares of size $k+2$.
    Let the strip width be $W = (2k+4)(k+3)$.
    An optimum packing of the instance is given in~\cref{fig:OPT_packing}(a).
    Notice that $2k+5$ squares of size $k+2$ together with two unit squares fit next to each other, because it holds that
    \begin{align*}
        (2k+5)(k+2) + 2 ~=~ 2k^2 + 9k + 12 ~\leq~ 2k^2+10k+12 ~=~ (2k+4)(k+3) ~=~ W.
    \end{align*}
    Moreover, since the $2k+4$ unit squares fit into two unit-width columns of height $k+2$, it follows that the height of the optimum packing is $h_{\text{OPT}} = k+2$.
    \begin{figure}[ht]
        \centering
        \begin{tikzpicture}[scale=0.3]
            \def\W{19.5};
            \draw[thick] (0,0) -- (\W,0) node[anchor=north west] {};
            \draw[thick,->] (0,0) -- (0,6.8) node[anchor=south east] {};
            \draw[thick,->] (\W,0) -- (\W,6.8) node[anchor=south east] {};

            \filldraw[myfill] (0,0) rectangle (6,6);
            \filldraw[myfill] (8,0) rectangle (14,6);

            \filldraw[myfill] (14,0) rectangle (15,1);
            \filldraw[myfill] (15,0) rectangle (16,1);
            \filldraw[myfill] (14,1) rectangle (15,2);
            \filldraw[myfill] (15,1) rectangle (16,2);
            \filldraw[myfill] (14,4) rectangle (15,5);
            \filldraw[myfill] (15,4) rectangle (16,5);
            \filldraw[myfill] (14,5) rectangle (15,6);
            \filldraw[myfill] (15,5) rectangle (16,6);

            \draw (3,3) node[] {$k+2$};
            \draw (7,3) -- (7,3) node {$\cdots$};
            \draw (11,3) node[] {$k+2$};
            \draw (15,3.2) -- (15,3.2) node {$\vdots$};

            \draw [thick, decorate, decoration = {calligraphic brace}] (14,-0.2) --  (0,-0.2);
            \draw (7,-1) node[] {$2k+5$};

            \draw [thick, decorate, decoration = {calligraphic brace}] (16.2,6) --  (16.2,0);
            \draw (17.9,3) node[] {$k+2$};

            \draw (\W * 0.6, -2.5) node {(a)};
        \end{tikzpicture} \hspace*{4mm}
        \begin{tikzpicture}[scale=0.3]
            \def\W{30};
            \draw[thick] (0,0) -- (\W,0) node[anchor=north west] {};
            \draw[thick,->] (0,0) -- (0,12.8) node[anchor=south east] {};
            \draw[thick,->] (\W,0) -- (\W,12.8) node[anchor=south east] {};

            \filldraw[myfill] (0,0) rectangle (1,1);
            \filldraw[myfill] (1,0) rectangle (7,6);
            \filldraw[myfill] (7,0) rectangle (8,1);
            \filldraw[myfill] (8,0) rectangle (14,6);
            \filldraw[myfill] (16,0) rectangle (17,1);
            \filldraw[myfill] (17,0) rectangle (23,6);
            \filldraw[myfill] (23,0) rectangle (24,1);
            \filldraw[myfill] (24,0) rectangle (30,6);

            \filldraw[myfill] (0,6) rectangle (6,12);

            \draw (4,3) node[] {$k+2$};
            \draw (11,3) node[] {$k+2$};
            \draw (15,3) -- (15,3) node {$\cdots$};

            \draw (20,3) node[] {$k+2$};
            \draw (27,3) node[] {$k+2$};

            \draw (3,9) node[] {$k+2$};

            \draw [thick, decorate, decoration = {calligraphic brace}] (\W,-0.2) --  (0,-0.2);
            \draw (\W * 0.5,-1) node[] {$2k+4$};

            \draw (\W * 0.4, -2.5) node {(b)};
        \end{tikzpicture}
        \vspace*{-2mm}
        \caption{(a) is an optimum packing of the instance.
        (b) is an optimal bottom-left $k$-local search packing of the instance. The unit and large square on the bottom of the strip repeat a total of $2k+4$ times.} 
        \label{fig:OPT_packing}
    \end{figure}
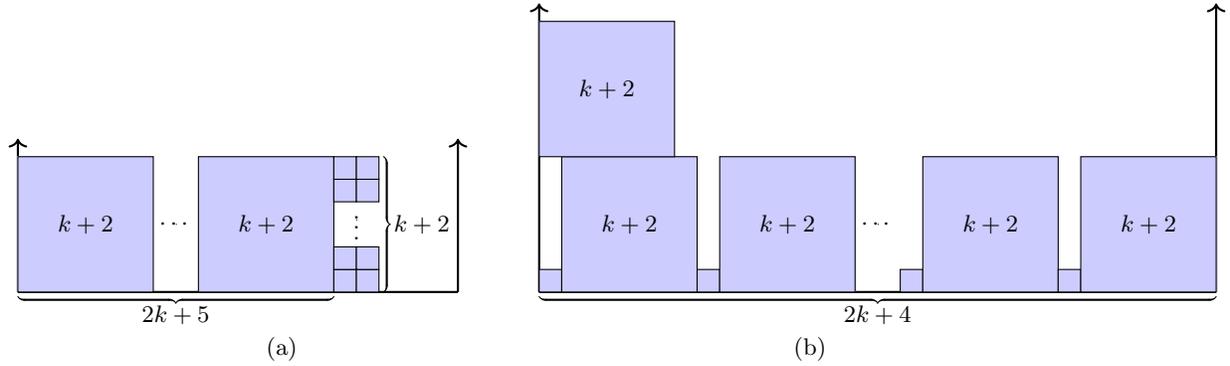

    Next, bottom-left place unit squares and big squares alternately, followed by another big square, as illustrated in \cref{fig:OPT_packing}(b).
    There does not exist a permutation of~$k$ squares such that the height of this bottom-left packing improves.
    First of all, even when~$k$ of the unit squares are removed, the big top square does not fit in the resulting gap,
    because
    \begin{align*}
        (2k+5) \cdot (k+2) + (2k+4 - k) \cdot 1 ~=~ 2k^2 + 10k + 14 ~>~ 2k^2 + 10k + 12 ~=~ W.
    \end{align*}
    Secondly, permuting $k$ large squares with unit squares amounts to swapping at most $k/2$ such pairs, in which case at most $2\cdot \frac{k}{2}+1=k+1$ unit squares are adjacent in the bottom row, thus the top square does not fit on top of those $k+1$ unit squares as its size is $k+2$.
    Therefore, the packing is $k$-local optimal and we have~$\hkbl = 2(k+2)$. 
    Now we get $\hkbl/\hopt = 2(k+2)/(k+2) = 2$ as claimed.
    \qed
\end{proof}

Moreover,  the bottom-left $k$-local search algorithm might need an exponential number of improvement steps until reaching a local optimum:

\begin{theorem}
    For each $k\in\mathbb{N}$, there exists a SPP-instance such that the bottom-left $k$-local search algorithm takes an exponential number of improvement steps with respect to the instance size.  
\end{theorem}
\begin{proof}
    Consider an instance consisting of rectangles $r_i=\left(\frac{1}{k},2^i\right)$ and $r_i'=\left(1,\frac{1}{k}\right)$ for~$i=0,\dots,k-1$.
    Let the strip width be~$1$.
    Order the rectangles as $r_0,r_0',r_1,r_1',\dots,r_{k-1},r_{k-1}'$.
    As depicted in~\cref{fig:exponential_many_improvement_steps}, the height of the bottom-left packing with regards to the initial ordering of the instance equals 
    \begin{align*}
        1 + \frac{1}{k} + 2 + \frac{1}{k} + 4 + \frac{1}{k} + \cdots + \frac{1}{k} + 2^{k-1} + \frac{1}{k} ~=~ 2^{k} - 1 + k\cdot \frac{1}{k} ~=~ 2^k.
    \end{align*}
    The rectangles $r_i$ are vertical and $r_i'$ are horizontal.
    There exists a sequence of $k$-permutations of the vertical rectangles such that the height of the bottom-left packing counts down from $2^k$ to $2^{k-1}+1$.
    Namely, after the $p$-th permutation the height of the bottom-left packing must be $2^k - p$.
    Write the number $2^k-p-1$ in its unique binary representation $\sum_{j=0}^{k-1} a_j 2^j$ with $a_j\in\{0,1\}$.
    Now permute the vertical rectangles such that~$2^j$ with $a_j = 1$ is placed before all~$2^\ell$ with $\ell<j$ and $a_\ell=0$ and after all~$2^\ell$ with $\ell>j$ and $a_\ell = 1$.
    Obviously, the height of this bottom-left packing is $2^k-p$, because the rectangles of size $2^\ell$ with $a_\ell=0$ fit into holes and do not account to the height of the packing, while all rectangles with $a_\ell=1$ do account to the height of the packing.
    As the height of the sequence counts down, it follows that the number of improvement steps is exponential in the input size.
\qed

    \begin{figure}[ht]
        \centering
        \vspace*{-4mm}
        \begin{tikzpicture}[scale=0.4]
            \draw[thick] (0,0) -- (2,0) node[anchor=north west] {};
            \draw[thick,->] (0,0) -- (0,17) node[anchor=south east] {};
            \draw[thick,->] (2,0) -- (2,17) node[anchor=south east] {};
            \foreach \y in {16}
                \draw (1pt,\y cm) -- (-1pt,\y cm) node[anchor=east] {$\y$};

            \filldraw[myfill] (0,0) rectangle (0.5,1);
            \filldraw[myfill] (0,1) rectangle (2,1.25); 
            
            \filldraw[myfill] (0,1.25) rectangle (0.5,3.25);
            \filldraw[myfill] (0,3.25) rectangle (2,3.5); 
            
            \filldraw[myfill] (0,3.5) rectangle (0.5,7.5);
            \filldraw[myfill] (0,7.5) rectangle (2,7.75); 
            
            \filldraw[myfill] (0,7.75) rectangle (0.5,15.75);
            \filldraw[myfill] (0,15.75) rectangle (2,16); 
        \end{tikzpicture}~~
        \begin{tikzpicture}[scale=0.4]
            \draw[thick] (0,0) -- (2,0) node[anchor=north west] {};
            \draw[thick,->] (0,0) -- (0,16) node[anchor=south east] {};
            \draw[thick,->] (2,0) -- (2,16) node[anchor=south east] {};
            \foreach \y in {15}
                \draw (1pt,\y cm) -- (-1pt,\y cm) node[anchor=east] {$\y$};

            \filldraw[myfill] (0,0) rectangle (0.5,2);
            \filldraw[myfill] (0,2) rectangle (2,2.25); 
            
            \filldraw[myfill] (0.5,0) rectangle (1,1);
            \filldraw[myfill] (0,2.25) rectangle (2,2.5); 

            \filldraw[myfill] (0,2.5) rectangle (0.5,6.5);
            \filldraw[myfill] (0,6.5) rectangle (2,6.75); 

            \filldraw[myfill] (0,6.75) rectangle (0.5,14.75);
            \filldraw[myfill] (0,14.75) rectangle (2,15); 
        \end{tikzpicture}~~
        \begin{tikzpicture}[scale=0.4]
            \draw[thick] (0,0) -- (2,0) node[anchor=north west] {};
            \draw[thick,->] (0,0) -- (0,15) node[anchor=south east] {};
            \draw[thick,->] (2,0) -- (2,15) node[anchor=south east] {};
            \foreach \y in {14}
                \draw (1pt,\y cm) -- (-1pt,\y cm) node[anchor=east] {$\y$};

            \filldraw[myfill] (0,0) rectangle (0.5,1);
            \filldraw[myfill] (0,1) rectangle (2,1.25); 

            \filldraw[myfill] (0,1.25) rectangle (0.5,5.25);
            \filldraw[myfill] (0,5.25) rectangle (2,5.5); 

            \filldraw[myfill] (0.5,1.25) rectangle (1,3.25);
            \filldraw[myfill] (0,5.5) rectangle (2,5.75); 
            
            \filldraw[myfill] (0,5.75) rectangle (0.5,13.75);
            \filldraw[myfill] (0,13.75) rectangle (2,14); 
        \end{tikzpicture}~~
        \begin{tikzpicture}[scale=0.4]

            \draw[thick] (0,0) -- (2,0) node[anchor=north west] {};
            \draw[thick,->] (0,0) -- (0,14) node[anchor=south east] {};
            \draw[thick,->] (2,0) -- (2,14) node[anchor=south east] {};
            \foreach \y in {13}
                \draw (1pt,\y cm) -- (-1pt,\y cm) node[anchor=east] {$\y$};

            \filldraw[myfill] (0,0) rectangle (0.5,4);
            \filldraw[myfill] (0,4) rectangle (2,4.25); 

            \filldraw[myfill] (0.5,0) rectangle (1,2);
            \filldraw[myfill] (0,4.25) rectangle (2,4.5); 
            
            \filldraw[myfill] (1,0) rectangle (1.5,1);
            \filldraw[myfill] (0,4.5) rectangle (2,4.75); 

            \filldraw[myfill] (0,4.75) rectangle (0.5,12.75);
            \filldraw[myfill] (0,12.75) rectangle (2,13); 
        \end{tikzpicture}~~
        \begin{tikzpicture}[scale=0.4]

            \draw[thick] (0,0) -- (2,0) node[anchor=north west] {};
            \draw[thick,->] (0,0) -- (0,13) node[anchor=south east] {};
            \draw[thick,->] (2,0) -- (2,13) node[anchor=south east] {};
            \foreach \y in {12}
                \draw (1pt,\y cm) -- (-1pt,\y cm) node[anchor=east] {$\y$};

            \filldraw[myfill] (0.5,3.5) rectangle (1,7.5);
            \filldraw[myfill] (0,11.75) rectangle (2,12); 

            \filldraw[myfill] (0,1.25) rectangle (0.5,3.25);
            \filldraw[myfill] (0,3.25) rectangle (2,3.5); 
            
            \filldraw[myfill] (0,0) rectangle (0.5,1);
            \filldraw[myfill] (0,1) rectangle (2,1.25); 

            \filldraw[myfill] (0,3.5) rectangle (0.5,11.5);
            \filldraw[myfill] (0,11.5) rectangle (2,11.75); 
        \end{tikzpicture}~~
        \begin{tikzpicture}[scale=0.4]
                
            \draw[thick] (0,0) -- (2,0) node[anchor=north west] {};
            \draw[thick,->] (0,0) -- (0,12) node[anchor=south east] {};
            \draw[thick,->] (2,0) -- (2,12) node[anchor=south east] {};
            \foreach \y in {11}
                \draw (1pt,\y cm) -- (-1pt,\y cm) node[anchor=east] {$\y$};

            \filldraw[myfill] (0.5,2.5) rectangle (1,6.5);
            \filldraw[myfill] (0,10.75) rectangle (2,11); 

            \filldraw[myfill] (0,0) rectangle (0.5,2);
            \filldraw[myfill] (0,2) rectangle (2,2.25); 
            
            \filldraw[myfill] (0.5,0) rectangle (1,1);
            \filldraw[myfill] (0,2.25) rectangle (2,2.5); 

            \filldraw[myfill] (0,2.5) rectangle (0.5,10.5);
            \filldraw[myfill] (0,10.5) rectangle (2,10.75); 
        \end{tikzpicture}~~
        \begin{tikzpicture}[scale=0.4]
                
            \draw[thick] (0,0) -- (2,0) node[anchor=north west] {};
            \draw[thick,->] (0,0) -- (0,11) node[anchor=south east] {};
            \draw[thick,->] (2,0) -- (2,11) node[anchor=south east] {};
            \foreach \y in {10}
                \draw (1pt,\y cm) -- (-1pt,\y cm) node[anchor=east] {$\y$};

            \filldraw[myfill] (0,1.25) rectangle (0.5,9.25);
            \filldraw[myfill] (0,9.25) rectangle (2,9.5); 

            \filldraw[myfill] (0.5,1.25) rectangle (1,5.25);
            \filldraw[myfill] (0,9.5) rectangle (2,9.75); 

            \filldraw[myfill] (1,1.25) rectangle (1.5,3.25);
            \filldraw[myfill] (0,9.75) rectangle (2,10); 
            
            \filldraw[myfill] (0,0) rectangle (0.5,1);
            \filldraw[myfill] (0,1) rectangle (2,1.25); 
        \end{tikzpicture}~~
        \begin{tikzpicture}[scale=0.4]
        
            \draw[thick] (0,0) -- (2,0) node[anchor=north west] {};
            \draw[thick,->] (0,0) -- (0,10) node[anchor=south east] {};
            \draw[thick,->] (2,0) -- (2,10) node[anchor=south east] {};
            \foreach \y in {9}
                \draw (1pt,\y cm) -- (-1pt,\y cm) node[anchor=east] {$\y$};

            \filldraw[myfill] (0,0) rectangle (0.5,8);
            \filldraw[myfill] (0,8) rectangle (2,8.25); 

            \filldraw[myfill] (0.5,0) rectangle (1,4);
            \filldraw[myfill] (0,8.25) rectangle (2,8.5); 

            \filldraw[myfill] (1,0) rectangle (1.5,2);
            \filldraw[myfill] (0,8.5) rectangle (2,8.75); 
            
            \filldraw[myfill] (1.5,0) rectangle (2,1);
            \filldraw[myfill] (0,8.75) rectangle (2,9); 
        \end{tikzpicture}
        \caption{Sequence of improvement steps of an instance in the bottom-left $k$-local search algorithm with $k=4$.}
        \label{fig:exponential_many_improvement_steps}
    \end{figure}
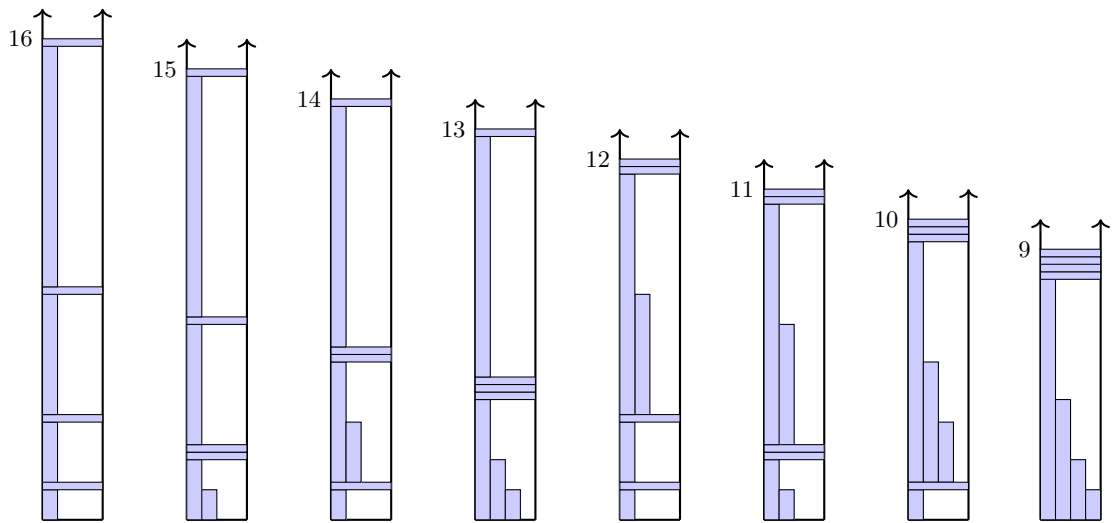

\end{proof}

It remains an open question whether the bottom-left $k$-local search performs better when the initial solution is guaranteed to be in some form.
For example, is the approximation ratio better-than-$2$ when the initial instance is ordered by decreasing width.
Another question is whether the $k$-local search algorithm takes a polynomial number of improvement steps when the algorithm always selects the best permutation.

%
%

 \bibliographystyle{splncs04}
 \bibliography{refs.bib}

\newpage
\section{Appendix}

In this appendix we present our proof for the upper bound stated in 
\cref{thm:worst_case_ordering_bounds}, i.e.\ we show that for every instance of the Square Strip Packing problem, the ratio between the height of the bottom-left packing using the worst ordering of the squares and the height of the optimal packing is smaller than a constant.
The remainder of this section is dedicated to proving that such a constant exists.
While it might be interesting to find a constant that is as small as possible, this will not be our objective.

The main strategy, to prove that the approximation ratio is bounded, is to cover the unoccupied space in a bottom-left packing of a SSPP-instance by a fixed number of copies of the squares.
This will imply that the ratio between the height of the bottom-left packing and the height of an optimal packing is bounded.
To find a covering of all the unoccupied space in a packing, it is first necessary to study the structure of the bottom-left packing.
More precisely, the relation between the relative positions of adjacent squares that form the boundary of unoccupied space is studied.

Instead of looking at a concrete packing, often the packing is replaced by an abstract representation of the packing in the form of a digraph.
Section~\ref{sec:adj_graph} defines this so-called adjacency graph which is the bedrock of the language for this section.
Next, the core result of Section~\ref{sec:bnd_structure_thm} is the structure theorem (Theorem~\ref{thm:boundedness_structure_thm}) describing the relative relation between squares that form the boundary of unoccupied space.
The structure theorem is used to construct a partition of unoccupied space.
The cover partition is the subject of Section~\ref{sec:cover_partition}.
In Section~\ref{sec:local_cov_thm}, the local cover theorem (Theorem~\ref{thm:local_cover_thm}) uses the cover partition together with the structure theorem to locally cover unoccupied space by  squares that form its boundary.
Section~\ref{sec:trenches} studies unoccupied space that is not bounded by squares and also describes local coverings of these so-called trenches.
It turns out that all these local coverings are compatible with each other, in the sense that there is no square that is used in arbitrarily many local coverings.
Hence, there exists a constant~$f$, such that the unoccupied space can be covered by at most~$f$ copies of the squares.
The compatibility of the local coverings is demonstrated in the global cover theorem (Theorem~\ref{thm:global_cover_thm}) in Section~\ref{sec:global_cov_thm}.
Finally, Section~\ref{sec:boundedness} brings everything together and shows that the approximation ratio~$h_{\text{BL}}^{\text{worst}}/h_{\text{OPT}}$ is bounded for SSPP-instances.

\subsection{The adjacency graph}\label{sec:adj_graph}
This section introduces the language of the structure theorem.
First, the unoccupied space is partitioned in different categories.
Next, the relation between the squares forming the boundary of unoccupied space is studied by defining the so-called adjacency graph.
This digraph induces different types of arrows.
These arrow types play a crucial role in the formulation of the structure theorem (Theorem~\ref{thm:boundedness_structure_thm}) in Section~\ref{sec:bnd_structure_thm}.

In easy terms, the unoccupied space is all the space in the substrip $[0,W]\times[0,h_{\text{BL}}]$ of a bottom-left packing that is not occupied by squares.
The following definition describes connected unoccupied space that gets bounded after placing the $i$-th square from the instance.
Although some parts of this unoccupied space might be filled by squares later, in the sequel it will be useful to have considered the unoccupied space and its surrounding squares at the moment it gets bounded.
\begin{definition}
    Let $\text{BL}\colon \mathcal{S}\to [0,W]\times[0,\infty)$ be the bottom-left packing of $\mathcal{S}$ in the order $\mathcal{S}_{\mathcal{A}} = (S_1,\dots,S_n)$.
    For $1\leq i\leq n$, define the $i$-subinstance $(\mathcal{S}_{\mathcal{A}})_i$ to be the subinstance of $\mathcal{S}_{\mathcal{A}}$ consisting of the first $i$ squares, that is $\mathcal{S}_{\mathcal{A}} = (S_1,\dots,S_i)$.
    Inductively define \textit{(unoccupied) $i$-pieces} to be the bounded connected maximal subspace~$U_1^i,\dots,U_{n_i}^i$ of
    \begin{align*}
        ([0,W]\times[0,\infty)) \setminus (\{\text{BL}(S) \mid S\in (\mathcal{S}_\mathcal{A})_{i}\}\cup\{U_j^k \mid 1\leq k<i,\ 1\leq j\leq n_k\}).
    \end{align*}
\end{definition}

Notice that $n_i$ is the number of $i$-pieces in the packing.
Often an unoccupied $i$-piece is just called an unoccupied piece if the value of $i$ is redundant. 
An unoccupied piece is an open polygon whose boundary consists of horizontal and vertical line segments, because a piece is bounded by squares that are closed subsets and whose faces are horizontal and vertical line segments.
This remark makes it possible to talk about the top, bottom, left and right face of an unoccupied piece.
\begin{definition}
    Define the \textit{top}, \textit{bottom}, \textit{left}, respectively \textit{right face} of a piece $U$ (or a square) by
    \begin{align*}
        \text{tf}(U) &~=~ \max\{ y \mid (x,y) \in U\}, \\
        \text{bf}(U) &~=~ \min\{ y \mid (x,y) \in U\}, \\
        \text{lf}(U) &~=~ \min\{ x \mid (x,y) \in U\}, \\
        \text{rf}(U) &~=~ \max\{ x \mid (x,y) \in U\}.
    \end{align*}
\end{definition}
Two unoccupied pieces are never adjacent, because pieces are maximal by definition.
Two squares are adjacent if their intersection is non-empty.
However, as squares are closed subsets and pieces are open subsets, this definition is not compatible to adjacency between a square and a piece.
Hence, in the next definition, to compare a piece and a square the closure of the piece is taken.
The closure of $U$ is denoted by $\overline{U}$.
\begin{definition}  
    Let $\sigma$ be an orthogonal packing of a SPP-instance $\mathcal{I}$.
    Then $S_1,S_2\in\mathcal{S}$ are \textit{adjacent} if it holds that $\sigma(S_1)\cap\sigma(S_2) \neq \varnothing$.
    And an unoccupied piece $U$ is \textit{adjacent} to $S_1$ if $\sigma(\overline{U}) \cap \sigma(S_1) \neq\varnothing$.
\end{definition}

Next, three different kind of unoccupied pieces are defined depending on the position of the piece in the strip.
\begin{definition}
    An unoccupied piece is called a \textit{left piece} if the piece is adjacent to the left strip boundary.
    It is a \textit{right piece} if the piece is adjacent to the right strip boundary.
    Otherwise a piece is called a \textit{middle piece}.
\end{definition}

The three classes of pieces partition the collection of all pieces.
Observe that a piece can never be a left piece and a right piece at the same time, because this would violate the bottom-left structure of the packing, as then there is a path contained in the piece from the left boundary to the right boundary, hence a square above the path could have been placed lower.
Thus the three classes are disjoint, hence forming a partition of the collection of unoccupied pieces.
There can also be unbounded unoccupied space in the packing of the strip $[0,W]\times[0,\infty)$.
This unbounded unoccupied space is the topic of Section~\ref{sec:trenches}.

Next, three formal squares are defined that represent the boundary.
The position of these formal squares is fixed for each packing.
The main reason to introduce formal squares is to be able to generalize a lot of statements in the upcoming sections, instead of having to distinguish between left, middle and right pieces.

\begin{definition}
    Let $\mathcal{I}=(\mathcal{S},W)$ be a SSPP-instance.
    Define the \textit{formal squares} $S_{\text{left}},S_{\text{right}}$ and~$S_{\text{bottom}}$ such that under each orthogonal packing $\sigma$ it holds that
\begin{align*}
        \sigma(S_{\text{left}}) &~=~ \{0\}\times \R_{\geq0}, \\
        \sigma(S_{\text{right}}) &~=~ \{W\}\times \R_{\geq0}, \\
        \sigma(S_{\text{bottom}}) &~=~ [0,W]\times \{0\}.
\end{align*}
    Define the \textit{formal instance of $\mathcal{I}$} to be $\widehat{\mathcal{I}}=(\widehat{\mathcal{S}},W)$ where $\widehat{\mathcal{S}} = \mathcal{S}\cup \{S_{\text{left}}, S_{\text{right}}, S_{\text{bottom}}\}$.
\end{definition}

\begin{definition}
    Consider a feasible packing of a (formal) SSPP-instance $\mathcal{I}= (\mathcal{S},W)$.
    Define the \textit{adjacency graph}~$G_{\text{adj}}(\mathcal{I})$ as the directed version of the graph with vertex set $\mathcal{S}$ and an edge between two vertices if the corresponding squares are adjacent.
\end{definition}

The adjacency graph is a connected planair digraph that describes all the adjacency relations in the packing of an instance.
On the contrary, the following definition only represents the adjacency relations between (formal) squares adjacent to an unoccupied piece.
This is the graph that will be used most throughout this section.

\begin{definition}
    Let $\mathcal{I}= (\mathcal{S},W)$ be an SSPP-instance and  $\widehat{\mathcal{I}_i}$ the formal instance of $i$-sub\-instance $\mathcal{I}_i$.
    Let~$U$ be an unoccupied $i$-piece and assume that $\mathcal{S}'\subseteq \widehat{\mathcal{S}}_i$ are the (formal) squares adjacent to $U$.
    Then define the \textit{adjacency graph of~$U$} to be the adjacency graph of $\widehat{\mathcal{I}_i}$ restricted to $\mathcal{S}'$, that is, 
    \begin{align*}
        G_{\text{adj}}(U) ~:=~ G_{\text{adj}}(\widehat{\mathcal{I}_i})[\mathcal{S}'].
    \end{align*}
\end{definition}
The adjacency graph can be seen as an undirected graph or as a digraph, because as digraph it has for every arrow another arrow going the opposite direction.
The main reason to talk about arrows is to define the four different types: up, down, right and left type.
\begin{definition}
    Let $(S_1,S_2)$ be an arrow in the adjacency graph $G_{\text{adj}}(\mathcal{I})$ of a SSPP-instance.
    \begin{enumerate}
        \item[(a)] $(S_1,S_2)$ is of \textit{left type} if $\text{lf}(S_1) = \text{rf}(S_2)$.
        \item[(b)] $(S_1,S_2)$ is of \textit{right type} if $\text{rf}(S_1) = \text{lf}(S_2)$.
        \item[(c)] $(S_1,S_2)$ is of \textit{up type} if $\text{tf}(S_1) = \text{bf}(S_2)$.
        \item[(d)] $(S_1,S_2)$ is of \textit{down type} if $\text{bf}(S_1) = \text{tf}(S_2)$.
    \end{enumerate}
\end{definition}

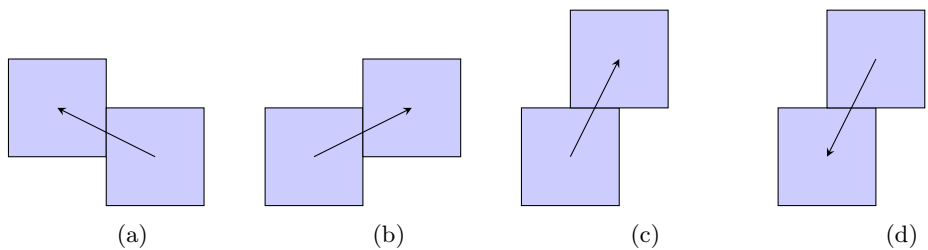
\begin{figure}[ht]
    \centering
    \begin{subfigure}{0.2\textwidth}
        \begin{tikzpicture}[scale=0.65]
            \filldraw[myfill] (0,1) rectangle (2,3);
            \filldraw[myfill] (2,0) rectangle (4,2);
            \draw [-stealth](3,1) -- (1,2); 
        \end{tikzpicture}
        \caption{}
        \label{fig:left_type}
    \end{subfigure}
    \begin{subfigure}{0.2\textwidth}
        \begin{tikzpicture}[scale=0.65]
            \filldraw[myfill] (0,0) rectangle (2,2);
            \filldraw[myfill] (2,1) rectangle (4,3);
            \draw [-stealth](1,1) -- (3,2); 
        \end{tikzpicture}
        \caption{}
        \label{fig:right_type}
    \end{subfigure}
    \begin{subfigure}{0.2\textwidth}
        \begin{tikzpicture}[scale=0.65]
            \filldraw[myfill] (0,0) rectangle (2,2);
            \filldraw[myfill] (1,2) rectangle (3,4);
            \draw [-stealth](1,1) -- (2,3); 
        \end{tikzpicture}
        \caption{}
        \label{fig:up_type}
    \end{subfigure}
    \begin{subfigure}{0.2\textwidth}
        \begin{tikzpicture}[scale=0.65]
            \filldraw[myfill] (0,0) rectangle (2,2);
            \filldraw[myfill] (1,2) rectangle (3,4);
            \draw [-stealth] (2,3) -- (1,1); 
        \end{tikzpicture}
        \caption{}
        \label{fig:down_type}
    \end{subfigure}
    \caption{Left type, right type, up type and down type respectively.}
    \label{fig:arrow_types}
\end{figure}

For an illustration of the different arrow types see \cref{fig:arrow_types}.
Furthermore, be cautious of the special case where two squares are only adjacent on a corner.
For example, let the bottom right corner of $S_1$ be adjacent to the top left corner of $S_2$, then $(S_1,S_2)$ is of right type as well as of down type.
Moreover, notice that the arrow from the formal square $S_{\text{bottom}}$ to $S_{\text{left}}$ is of left type as well as of up type.
Similarly, the arrow from~$S_{\text{bottom}}$ to $S_{\text{right}}$ is of right type as well as of up type.

\subsection{The structure theorem}\label{sec:bnd_structure_thm}
The structure theorem (Theorem~\ref{thm:boundedness_structure_thm}) describes the relative position of adjacent squares in the adjacency graph of an unoccupied piece.
In more detail, this section introduces a few special squares, among which are the start square and end square.
There are two paths in the adjacency graph from the start square to the end square, the top path going over the unoccupied piece and the bottom path going under the unoccupied piece.
Essentially, the structure theorem describes the different types of arrows that occur in the top and bottom path.
In Section~\ref{sec:cover_partition} and~\ref{sec:local_cov_thm} the structure theorem will be exploited to construct a covering of the unoccupied space in a bottom-left packing.

The adjacency graph of an unoccupied piece contains a closed walk surrounding the piece, because a piece is bounded by definition and the adjacency graph contains the formal squares, hence every point on the boundary of a piece is part of a (formal) square.
Actually, Lemma~\ref{lem:hamcircuit} will show something stronger, namely, that there exists  a Hamiltonian circuit surrounding a piece in the adjacency graph of the piece.
This circuit will be used to define the top and bottom paths that are required in the formulation of the structure theorem.
The main idea for the proof of Lemma~\ref{lem:hamcircuit} is to use the natural ordering induced by the boundary of the unoccupied piece and show that this closed walk is already a Hamiltonian circuit.
\begin{lemma}\label{lem:hamcircuit}
    Let $U$ be an unoccupied piece.
    There exists a Hamiltonian circuit in the adjacency graph of~$U$ that surrounds~$U$.
\end{lemma}
\begin{proof}
    For each square adjacent to~$U$ there is at least one point on the boundary~$\partial U$ of the piece that intersects the square and for each point on the boundary there are at most three squares intersecting this point.
    The boundary $\partial U$  is homeomorphic to a circle, thus an orientation of the circle naturally gives an ordering of the squares adjacent to~$U$.
    Hence this induces a closed walk~$W$ in the adjacency graph of~$U$ that surrounds~$U$ and visits each vertex at least once.

    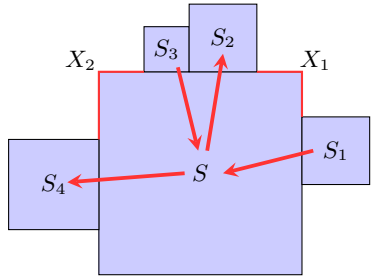
\begin{figure}[ht]
    \centering
    \begin{tikzpicture}[scale=0.3]
        \filldraw[myfill] (0,0) rectangle (9,9);
        \node at (4.5,4.5) {$S$};        

        \filldraw[myfill] (9,4) rectangle (12,7);
        \node at (10.5,5.5) {$S_1$};        

        \filldraw[myfill] (2,9) rectangle (4,11);
        \node at (3,10) {$S_3$};        
        \filldraw[myfill] (4,9) rectangle (7,12);
        \node at (5.5,10.5) {$S_2$};        
        
        \filldraw[myfill] (-4,2) rectangle (0,6);
        \node at (-2,4) {$S_4$};  

        \draw [-stealth, line width=0.5mm, red!80!white] (9.5,5.5) -- (5.5,4.5);
        \draw [-stealth, line width=0.5mm, red!80!white] (4.8,5.5) -- (5.5,9.8);
        \draw [-stealth, line width=0.5mm, red!80!white] (3.5,9.2) -- (4.4,5.5);
        \draw [-stealth, line width=0.5mm, red!80!white] (3.8,4.5) -- (-1.4,4.1);

        \draw[thick, red!80!white] (7,9) -- (9,9) -- (9,7);
        \node at (9.6,9.5) {$X_1$};  
        \draw[thick, red!80!white] (0,6) -- (0,9) -- (2,9);
        \node at (-0.8,9.5) {$X_2$};  
    \end{tikzpicture}
    \caption{The arrows $(S_1,S),(S,S_2), (S_3,S)$ and $(S,S_4)$ together with $X_1$ and $X_2$.
    In this example, $F_1$ is either the right or top face of~$S$ and $F_2$ is either the left or top face of~$S$.}
    \label{fig:X1andX2}
    \end{figure}

    \textbf{Claim}: $W$ is a Hamiltonian circuit, that is, $W$ visits each square exactly once.
    This claim is proven by contradiction.
    Suppose that there is a square~$S$ that is visited twice.
    Then there are four arrows $(S_1,S)$, $(S,S_2)$, $(S_3,S)$ and $(S,S_4)$ such that the path from $S_4$ to $S_1$ surrounds the piece.
    This is illustrated in \cref{fig:X1andX2}.
    Let $X_1\subseteq \partial S$ be the part of the boundary of $S$ between $S_1$ and $S_2$ that is adjacent to $U$.
    Similar, let $X_2\subseteq \partial S$ be that between $S_3$ and~$S_4$ adjacent to $U$.
    Let $F_1$ and $F_2$ be faces of $S$ such that $F_1\cap X_1$ respectively $F_2\cap X_2$ do not have measure zero as subset of $F_1$ or $F_2$.
    Distinguish the following cases depending on the position of $F_1$ and $F_2$ relative to $S$.

    \textbf{Case 1}: Let $F_1$ and $F_2$ be the same face of $S$.
    Regardless of which face $F_1$ is, there is a square on the path between $S_2$ and $S_3$ that can be placed more bottom-left, because exactly one of the sets $X_1$ or~$X_2$ is either left or below of $S_2$ and $S_3$.

    \textbf{Case 2}: Let $F_1$ and $F_2$ be opposite faces of $S$.    
    Then $X_1$ and $X_2$ are on the same side of $S_2$ respectively~$S_3$, that is, either both $X_1$ below $S_2$ and $X_2$ below $S_3$, etc.
    In each case, there is a square on the path between~$S_2$ and~$S_3$ that can be placed more bottom-left. 

    \textbf{Case 3}: Let $F_1$ and $F_2$ be two adjacent faces of $S$. 
    If $F_1$ and $F_2$ are the left and bottom face of~$S$, then the path between $S_2$ and $S_3$ can be placed more bottom-left, as this path does not surround~$U$.
    Otherwise, at least one of the sets $X_1$ or $X_2$ is left or below $S_2$ and $S_3$, hence also a square on the path between $S_2$ and~$S_3$ can be placed more bottom-left.
    \qed
\end{proof}

The Hamiltonian circuit from Lemma~\ref{lem:hamcircuit} will be split into two directed paths, called the top path and the bottom path.
These two paths start in the start square and end in the end square.
The next definition defines these squares.
Additionally, a few other special squares are introduced that are relevant in the subsequent sections.
\begin{definition}
    Let $U$ be an unoccupied piece and let $W$ be the Hamiltonian circuit in the adjacency graph of~$U$ that surrounds $U$.
    Assume that $W$ is clockwise oriented.
    \begin{enumerate}
        \item[(a)] The \textit{start square} $S_{\text{start}}$ is a square corresponding to a vertex in the adjacency graph of~$U$ whose top face is as low as possible and among those squares it is the left most.
        \item[(b)] The \textit{top square} $S_{\text{top}}$ is a square corresponding to a vertex in the adjacency graph of~$U$ whose bottom face is as high as possible and among those squares it is the right most.
        \item[(c)] The \textit{end square} $S_{\text{end}}$ is the square after $S_{\text{top}}$ on the oriented Hamiltonian circuit~$W$.
        \item[(d)] The \textit{pre-top square} $S_{\text{pre}}$ is the square before $S_{\text{top}}$ on~$W$.
        \item[(e)] The \textit{penultimate square} $S_{\text{pen}}$ is the square after $S_{\text{end}}$ on~$W$.
    \end{enumerate}
\end{definition}

The adjacency graph of an unoccupied piece always has at least four vertices, because the boundary of an unoccupied piece consists of horizontal and vertical line segments, hence a rectangle is the smallest polygon enclosing unoccupied space.
It follows that the start square and the top square cannot be adjacent in the adjacency graph, as then the lowest top face and the highest bottom face are on the same height, in which case the piece is empty.
Consequently, the start, top, end and pre-top square are always different squares.
Contrarily, the penultimate square might be the same as the start square.

\begin{figure}[ht]
    \centering
    \begin{tikzpicture}[scale=0.3]
        \filldraw[myfill] (0,0) rectangle (1,1);            
        \filldraw[myfill] (1,0) rectangle (4,3);
        \filldraw[myfill] (4,2) rectangle (9,7);
        \filldraw[myfill] (9,6) rectangle (17,14);
        \filldraw[myfill] (9,2) rectangle (13,6);
        \filldraw[myfill] (12,0) rectangle (14,2);
        \filldraw[myfill] (14,0) rectangle (15,1);
        \filldraw[myfill] (15,0) rectangle (20,5);
        \filldraw[myfill] (20,4) rectangle (26,10);
        \filldraw[myfill] (19,10) rectangle (29,20);

        \draw [-stealth, line width=0.5mm, red!80!white] (0.5,0.5) -- (2.5,1.5);
        \draw [-stealth, line width=0.5mm, red!80!white] (2.5,1.5) -- (6.5,4.5);
        \draw [-stealth, line width=0.5mm, red!80!white] (6.5,4.5) -- (13,10);
        \draw [-stealth, line width=0.5mm, red!80!white] (13,10) -- (11,4);
        \draw [-stealth, line width=0.5mm, red!80!white] (11,4) -- (13,1);
        \draw [-stealth, line width=0.5mm, red!80!white] (13,1) -- (14.5,0.5);
        \draw [-stealth, line width=0.5mm, red!80!white] (14.5,0.5) -- (17.5,2.5);
        \draw [-stealth, line width=0.5mm, red!80!white] (17.5,2.5) -- (23,7); 
        \draw [-stealth, line width=0.5mm, red!80!white] (23,7) -- (24,15);
        
        \filldraw[myfill] (-4,0) rectangle (0,4);
        \filldraw[myfill] (-6,4) rectangle (1,11);
        \filldraw[myfill] (-10,4) rectangle (-6,8);
        \filldraw[myfill] (-18,8) rectangle (-9,17);

        \draw [myfill] (-21,21) -- (-21,17) -- (19,17) -- (19,21);

        \draw [-stealth, line width=0.5mm, red!80!white] (0.5,0.5) -- (-2,2);
        \draw [-stealth, line width=0.5mm, red!80!white] (-2,2) -- (-2.5,7.5);
        \draw [-stealth, line width=0.5mm, red!80!white] (-2.5,7.5) -- (-8,6);
        \draw [-stealth, line width=0.5mm, red!80!white] (-8,6) -- (-13.5,12);
        \draw [-stealth, line width=0.5mm, red!80!white] (-13.5,12) -- (0,19.5);
        \draw [-stealth, line width=0.5mm, red!80!white] (0,19.5) -- (24,15);

        \node[anchor=north] at (0.5,0) {$S_{\text{start}}$};
        \node[anchor=north] at (-3.75,20.7) {$S_{\text{top}}$};
        \node[anchor=north] at (-15,13.2) {$S_{\text{pre}}$};
        \node[anchor=north] at (25.8,16) {$S_{\text{end}}$};
        \node[anchor=north] at (24.5,7.8) {$S_{\text{pen}}$};

        \node[anchor=north] at (4,13) {$U$};        
    \end{tikzpicture}
    \caption{Example of an unoccupied piece~$U$ together with the top and bottom path in the adjacency graph.
    Only the squares adjacent to the piece are depicted, however, this particular example does exist in a bottom-left packing of squares.
    The top square is drawn as a rectangle for the sake of saving space.}
    \label{fig:top_bottom_paths}
\end{figure}
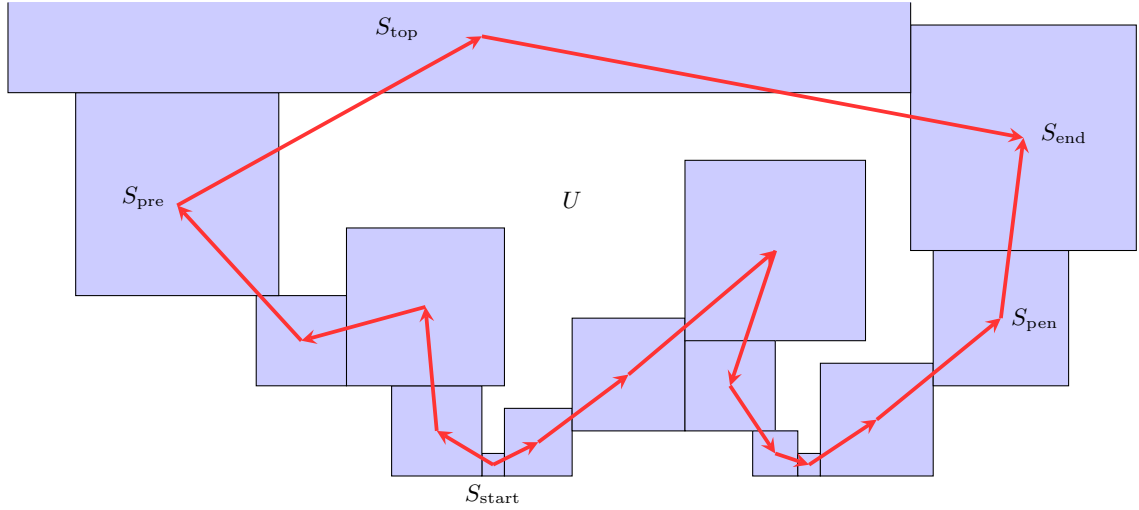

The next definition splits the Hamiltonian circuit from Lemma~\ref{lem:hamcircuit} into two paths between the start square and the end square.
This is also illustrated in \cref{fig:top_bottom_paths}.
\begin{definition}
    Let $U$ be an unoccupied piece and let $W$ be the Hamiltonian circuit in the adjacency graph of~$U$ that surrounds $U$.
    The \textit{top path} is the directed path in~$W$ from the start square to the end square traversing the top square.
    The \textit{bottom path} is the directed path in~$W$ from the start square to the end square not traversing the top square.
\end{definition}
Observe that the unoccupied piece is always to the right of the top path when traversed from start to end square.
Analoguously, the unoccupied piece is to the left of the bottom path.
Now, the structure theorem describes the types of the arrows in the top and bottom path.
The structure theorem is true for left, middle and right pieces.
\begin{theorem}[Structure theorem]\label{thm:boundedness_structure_thm}
    Let $U$ be an unoccupied piece.
    Then
    \begin{enumerate}
        \item[(a)] All arrows in the top path between the start and pre-top square are either of left type or of up type.

        \item[(b)] The arrow from the pre-top square to the top square is either of right type or of up type.
        \item[(c)] The arrow from the top square to the end square is either of right type or of down type.
    
        \item[(d)] Each arrow in the bottom path between the start square and the penultimate square is either of right type or of down type.
        \item[(e)] The arrow from the penultimate square to the end square is either of right or up type.
    \end{enumerate}
\end{theorem}
\begin{proof}
    Let $S_0,\dots,S_\ell$ be the top path and let $S_{\ell},S_{\ell+1},\dots,S_{\ell+r}$ be the bottom path in reversed order.
    That is, $S_\text{start} = S_0 = S_{\ell+r}$, $S_{\text{pre}} = S_{\ell-2}$, $S_{\text{top}} = S_{\ell-1}$, $S_{\text{end}} = S_{\ell}$ and $S_{\text{pen}} = S_{\ell+1}$.
    The unoccupied piece is always on the right of the oriented circuit~$S_{0},\dots,S_{\ell+r}$.

    Consider the arrows $(S_\text{start},S_1)$ and $(S_{\text{start}}, S_{\ell+r-1})$.
    Neither $(S_\text{start},S_1)$ nor $(S_{\text{start}}, S_{\ell+r-1})$ can be of down type, as then $S_{1}$ or $S_{\ell+r-1}$ would be below $S_{\text{start}}$, contradicting that $S_{\text{start}}$ is the lowest square adjacent to the unoccupied piece.
    Also, not both $(S_\text{start},S_1)$ and $(S_{\text{start}}, S_{\ell+r-1})$ can be of left type, as then at least one of $\text{tf}(S_1)$ and $\text{tf}(S_{\ell+r-1})$ is below $\text{tf}(S_{\text{start}})$.
    With the same reasoning, not both $(S_\text{start},S_1)$ and $(S_{\text{start}}, S_{\ell+r-1})$ can be of right type.
    Hence, as $(S_\text{start},S_1)$ is on the top path, it is of left or up type and as $(S_\text{start},S_{\ell+r-1})$ is on the bottom path, it is of right or up type.

    Next, consider the first arrow $(S_i,S_{i+1})$ on the circuit that is not of left or up type.
    Such an arrow exists, because if all arrows are of left and up type, then the circuit never returns to the start square.
    Also it holds that $i\geq 1$, because $(S_0,S_1)$ is of left or up type by the previous paragraph.
    Hence it is always possible to consider the square~$S_{i-1}$.

    \textbf{Case 1}: Suppose $(S_i,S_{i+1})$ is of down type.
    Then $(S_{i-1},S_i)$ cannot be of left type, as otherwise $S_i$ can be placed more to the left, because $S_{i-1}$ must be above $S_{i+1}$ and the unoccupied piece is to the right-hand side of the path $S_{i-1},S_i,S_{i+1}$.
    Thus $(S_{i-1},S_i)$ must be of up type.
    Now as $(S_i,S_{i+1})$ is of down type, it follows that each arrow $(S_j,S_{j+1})$ with $j>i$ must be of left or up type.
    Otherwise, let $(S_j,S_{j+1})$ with $j>i$ be the first that is not of left or up type.
    Suppose~$(S_j,S_{j+1})$ is of right type, then $S_j$ could have been placed lower.
    And suppose~$(S_j,S_{j+1})$ is of down type, then $S_j$ could have been placed more to the left.
    Thus it follows that~$S_i$ must be the top square, i.e., $i = \ell-1$.
    
    \textbf{Case 2}: Suppose that $(S_i,S_{i+1})$ is of right type. 
    Then $(S_{i+1},S_{i+2})$ must be of down type as otherwise~$S_{i+1}$ could have been placed lower.
    Now with the same argument as before, all the arrows $(S_j,S_{j+1})$ with $j>i+1$ must be of left or up type.
    Thus either $S_i$ or $S_{i+1}$ is the top square depending on which one has a bottom face that is higher.
    Thus this implies (a).

    If $(S_{\ell-1},S_{\ell})$ is of down type, then $(S_{\ell-2},S_{\ell-1})$ is of up type as mentioned in Case~1, this implies~(b) and~(c).
    If $(S_{i},S_{i+1})$ is the first arrow of right type, then there are two cases.
    First, if $\text{bf}(S_{i})>\text{bf}(S_{i+1})$, then $S_{i}$ is the top square.
    It must hold that $(S_{i-1},S_i)$ is of up type as otherwise $S_i$ could have been placed lower, this implies~(b) and~(c).
    Secondly, if $\text{bf}(S_{i})\leq \text{bf}(S_{i+1})$, then $S_{i+1}$ is the top square, this immediately implies~(b), and it follows from Case~2 that also~(c) is true.

    All arrows on the path $S_{\ell+1},\dots,S_{\ell+r}$ are of left or up type according to the argumentation above.
    Thus each arrow in the bottom path between the start square  $S_{\ell+r}$ and the penultimate square~$S_{\ell+1}$ is of right or down type, this implies~(d).
    If $(S_{\ell-1},S_{\ell})$ is of down type, then~$(S_{\ell},S_{\ell+1})$ is of left type, as else $S_{\ell}$ could have been placed more to the left.
    This implies~(e).
    Last of all, if $(S_{\ell-1},S_{\ell})$ is of right type, then $(S_{\ell},S_{\ell+1})$ must be of down type, implying~(e).
    \qed
\end{proof}

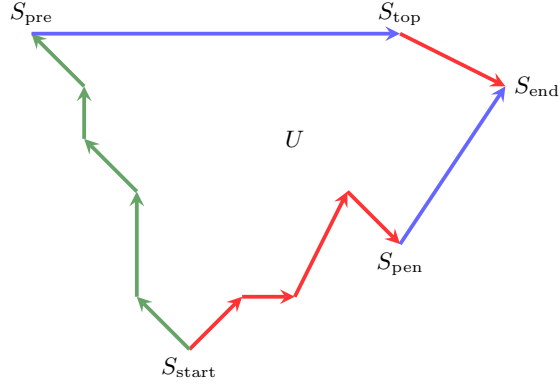
\begin{figure}[ht]
    \centering
    \begin{tikzpicture}[scale=0.7]
        \draw [-stealth, line width=0.5mm, red!80!white] (0,0) -- (1,1); 
        \draw [-stealth, line width=0.5mm, red!80!white] (1,1) -- (2,1);
        \draw [-stealth, line width=0.5mm, red!80!white] (2,1) -- (3,3);
        \draw [-stealth, line width=0.5mm, red!80!white] (3,3) -- (4,2);
        \draw [-stealth, line width=0.5mm, blue!60!white] (4,2) -- (6,5);

        \draw [-stealth, line width=0.5mm, black!60!green!60!white] (0,0) -- (-1,1);
        \draw [-stealth, line width=0.5mm, black!60!green!60!white] (-1,1) -- (-1,3);
        \draw [-stealth, line width=0.5mm, black!60!green!60!white] (-1,3) -- (-2,4);
        \draw [-stealth, line width=0.5mm, black!60!green!60!white] (-2,4) -- (-2,5);
        \draw [-stealth, line width=0.5mm, black!60!green!60!white] (-2,5) -- (-3,6);
        \draw [-stealth, line width=0.5mm, blue!60!white] (-3,6) -- (4,6);
        \draw [-stealth, line width=0.5mm, red!80!white] (4,6) -- (6,5);

        \node[] at (2,4) {$U$};
        \node[anchor=north] at (0,0) {$S_{\text{start}}$};
        \node[anchor=west] at (6,5) {$S_{\text{end}}$};
        \node[anchor=south] at (4,6) {$S_{\text{top}}$};
        \node[anchor=south] at (-3,6) {$S_{\text{pre}}$};
        \node[anchor=north] at (4,2) {$S_{\text{pen}}$};
    \end{tikzpicture}
    
    \caption{Example of the top and bottom path around a piece $U$.
    The green arrows are either of left or up type, the blue arrows are either of right or up type, and the red arrows are either of right or down type.}
    \label{fig:all_types}
\end{figure}

In line with the structure theorem, an illustration of the different types of arrows in the top and bottom path is given in \cref{fig:all_types}.
The next section studies the consequences of the structure theorem and applies it to find a covering of the unoccupied space by squares.

\subsection{Cover partitions}\label{sec:cover_partition} 
The main objective of this section is to divide an unoccupied piece into easily coverable parts called subpieces.
The idea of the so-called cover partition is that the width of each horizontal line in a subpiece increases with the height of the position of the horizontal line.
Hence, if a square is wider than the top of a subpiece, then it is wider than any horizontal line in the subpiece.
This will make it possible to cover the subpiece by the squares around it when the height of the subpiece is bounded by its width. 

This section starts by defining the line space, this makes it possible to talk about horizontal lines in an unoccupied piece.
After that, the cover partition is defined formally.
The line space is the set of all horizontal lines in an unoccupied piece.
\begin{definition}
    Let $V$ be a subspace of an unoccupied piece $U$.
    The \textit{(horizontal) line space of $V$} is defined as the set
    \begin{align*}
       \mathcal{L}V ~=~ \{\text{connected components of } \overline{V}\cap \ell_y\mid \ell_y = [0,W]\times \{y\},\ y\in [0,h_{\text{BL}}]\}.
    \end{align*}
    Here $\overline{V}$ denotes the closure of $V$.
\end{definition}

To compare lines in the line space with each other, a projection map is defined.
Moreover, the width of a line is defined in the obvious way.
\begin{definition}
    Let $V$ be a subspace of an unoccupied piece $U$.
    The \textit{projection} of horizontal lines in $V$ is the map
    \begin{align*}
        \phi \colon \mathcal{L}V \to 2^{[0,W]} : \ell=[x_0,x_1]\times\{y\} \mapsto [x_0,x_1].
    \end{align*}
    The \textit{width} of a horizontal line in $V$ is defined as
    \begin{align*}
        w\colon \mathcal{L}V\to \R_{\geq0}: \ell=[x_0,x_1]\times\{y\} \mapsto x_1 - x_0.
    \end{align*}
\end{definition}

With these tools, the cover partition of an unoccupied piece is introduced to be the smallest partition of the piece into connected subpieces, such that a pair of projections of horizontal lines is either included in one another depending on which horizontal line is higher, or their intersection is empty.
\begin{definition}\label{def:cover_partition}
    Let $U$ be an unoccupied piece.
    A \textit{cover partition} $\mathcal{P}_{\text{cov}}(U) = \{V_1,\dots,V_s\}$ is a minimal partition of $U$ such that
    \begin{enumerate}
        \item[(1)] For every $1\leq i\leq s$, the subspace $V_i$ of $U$ is connected.
        \item[(2)] For every $1\leq i\leq s$ and every pair of horizontal lines $\ell,\ell' \in \mathcal{L}V_i$ with $\ell$ below $\ell'$, it holds that either $\phi(\ell) \subseteq \phi(\ell')$ or $\phi(\ell)\cap \phi(\ell')=\varnothing$.
    \end{enumerate}
    Call the sets $V_1,\dots,V_s$ the \textit{subpieces} of~$U$.
    Here minimal means that the cover partition is not the refinement of another partition satisfying~(1) and (2).
\end{definition}

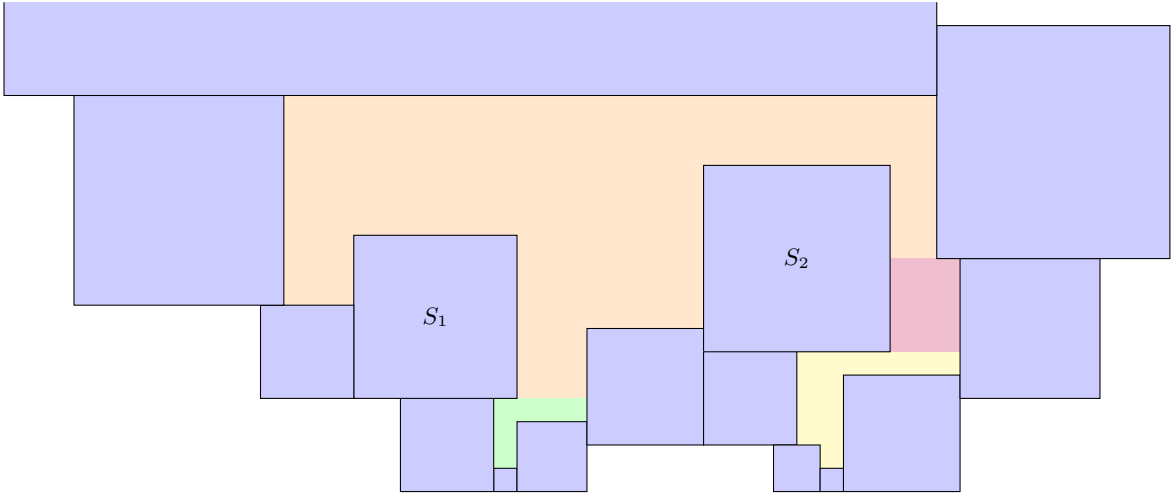
\begin{figure}[ht]
    \centering
    \begin{tikzpicture}[scale=0.31]
        \filldraw[bart_orange] (17,10) rectangle (19,17);
        \filldraw[bart_orange] (-9,14) rectangle (17,17);
        \filldraw[bart_orange] (-9,11) rectangle (9,14);
        \filldraw[bart_orange] (-9,4) rectangle (9,11);

        \filldraw[bart_green] (0,1) rectangle (1,4);
        \filldraw[bart_green] (1,3) rectangle (4,4);

        \filldraw[bart_yellow] (13,2) rectangle (14,6);
        \filldraw[bart_yellow] (14,1) rectangle (15,6);
        \filldraw[bart_yellow] (15,5) rectangle (20,6);
        
        \filldraw[bart_purple] (17,6) rectangle (20,10);

        \filldraw[myfill] (0,0) rectangle (1,1);            
        \filldraw[myfill] (1,0) rectangle (4,3);
        \filldraw[myfill] (4,2) rectangle (9,7);
        \filldraw[myfill] (9,6) rectangle (17,14);
        \filldraw[myfill] (9,2) rectangle (13,6);
        \filldraw[myfill] (12,0) rectangle (14,2);
        \filldraw[myfill] (14,0) rectangle (15,1);
        \filldraw[myfill] (15,0) rectangle (20,5);
        \filldraw[myfill] (20,4) rectangle (26,10);
        \filldraw[myfill] (19,10) rectangle (29,20);

        \filldraw[myfill] (-4,0) rectangle (0,4);
        \filldraw[myfill] (-6,4) rectangle (1,11);
        \filldraw[myfill] (-10,4) rectangle (-6,8);
        \filldraw[myfill] (-18,8) rectangle (-9,17);
        \draw [myfill] (-21,21) -- (-21,17) -- (19,17) -- (19,21);     

        \node[] at (-2.5,7.5) {$S_{1}$};
        \node[] at (13,10) {$S_{2}$};
        
    \end{tikzpicture}
    \caption{Example of the natural cover partition of an unoccupied piece.
    There are four subpieces indicated by the green, yellow, purple and orange area.
    Green is~$V_1$, yellow is~$V_2$, purple is~$V_{\text{end}}$ and orange is~$V_{\text{top}}$.
    The top square is drawn as a rectangle for the sake of convenience.}
    \label{fig:natural_cover_partition_example}
\end{figure}

\cref{fig:natural_cover_partition_example} depicts an example of a cover partition.
A cover partition of an unoccupied piece is not necessarily unique.
An explicit construction of a cover partition is given in Theorem~\ref{thm:existence_cover_partition}.
This construction will use that there is only one so-called peak in the adjacency graph of an unoccupied piece.
Intuitively, a peak square is a local top square, as shown in \cref{fig:peak_square}.
\begin{definition}
    Let $S_0,\dots,S_\ell,S_{\ell+1},\dots,S_{\ell+r}$ be the top path followed by the reversed bottom path in the adjacency graph of a piece $U$.
    For a vertex $S_i$, let $S_k$ be the vertex with $k<i$ maximal such that $\text{bf}({S_k}) \neq~\text{bf}({S_i})$ and let $S_j$ be the vertex with $j>i$ minimal such that $\text{bf}({S_j}) \neq \text{bf}({S_i})$.
    Then $S_i$ is a \textit{peak square} if it holds that $\text{bf}({S_k}) < \text{bf}({S_i})$ and $\text{bf}({S_j}) < \text{bf}({S_i})$.
\end{definition}
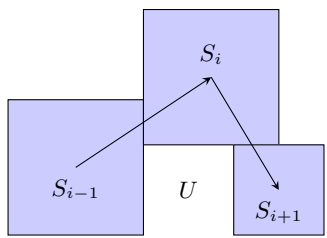
\begin{figure}[ht]
    \centering
    \begin{tikzpicture}[scale=0.6]
        \filldraw[myfill] (0,0) rectangle (3,3);
        \filldraw[myfill] (3,2) rectangle (6,5);
        \filldraw[myfill] (5,0) rectangle (7,2);
        \draw [-stealth] (1.5,1.5) -- (4.5,3.5);
        \draw [-stealth] (4.5,3.5) -- (6,1);
        \node[] at (4,1) {$U$};
        \node[] at (1.5,1) {$S_{i-1}$};
        \node[] at (4.5,4) {$S_i$};
        \node[] at (6,0.5) {$S_{i+1}$};
    \end{tikzpicture}
    \caption{Example of a peak square $S_i$.}
    \label{fig:peak_square}
\end{figure}

There can be at most two peak squares next to each other, as otherwise a peak square in the middle can be placed lower, contradicting the bottom-left placement rules.
Evidently, the top square is a peak square, because it is the square adjacent to $U$ with highest bottom face.
Moreover, Lemma~\ref{lem:peak_squares} shows that the structure theorem implies that there can be at most two peak squares, namely the top square, and possibly the pre-top square.
This will be useful for explicitly constructing a cover partition.
\begin{lemma}\label{lem:peak_squares}
    An unoccupied piece~$U$ has at most two peak squares, which are the top square and possibly the pre-top square.
\end{lemma}
\begin{proof}
    Let $S_0,\dots,S_\ell,S_{\ell+1},\dots,S_{\ell+r}$ be the top path followed by the reversed bottom path of~$U$, such that~$S_{\ell-1}$ is the top square.
    Consider a peak square~$S_i$.
    Let~$S_k$ with~$k<i$ be the vertex closest to~$S_i$ such that $\text{bf}(S_{k}) \neq \text{bf}(S_{i})$.
    And similarly, let~$S_j$ with~$j>i$ be the vertex closest to $S_i$ such that $\text{bf}(S_{j}) \neq \text{bf}(S_{i})$.
    Then the bottom face of~$S_i$ is on the same height as the bottom face of all the squares~$S_n$ with~$k<n<j$.
    Therefore, all the arrows $(S_n,S_{n+1})$ with $k<n<j-1$ are of right type.
    According to the structure theorem (Theorem~\ref{thm:boundedness_structure_thm}), only the arrow from the pre-top square to the top square, and the arrow from the top square to the end square can be of right type.
    However, notice that $\text{bf}(S_{\text{end}}) \neq \text{bf}(S_{\text{top}})$, as the end square is more to the right than the top square, hence contradicting the definition of the top square.
    Thus for all~$i\notin \{\ell-2,\ell-1\}$, with $S_i$ a peak square, it follows that $S_k = S_{i-1}$ and $S_j = S_{i+1}$.
    
    Now $(S_{i-1},S_i)$ is either of right or up type and $(S_{i},S_{i+1})$ is either of right or down type, otherwise there is no space on the bottom face of $S_i$ between $S_{i-1}$ and $S_{i+1}$ that is adjacent to~$U$.
    Thus there are four combinations.
    First of all, let both arrows be of right type, this never happens as then $S_i$ could have been placed lower.
    Secondly, let $(S_{i-1},S_i)$ be of right type and~$(S_i,S_{i+1})$ be of down type, then by the structure theorem $S_i$ is the top square or the end square.
    However, the end square cannot be a peak square as the bottom face of the top square is always above the bottom face of the end square, thus in this case $S_i$ is the top square.
    Thirdly, let~$(S_{i-1},S_i)$ be of up type and~$(S_i,S_{i+1})$ be of right type, then by the structure theorem $S_i$ is the pre-top square or the top square.
    Last of all, let~$(S_{i-1},S_i)$ be of up type and $(S_i,S_{i+1})$ be of down type, then $S_i$ is the top square according to the structure theorem.
    
    All in all, the pre-top and the top are the only possible peak squares.
    Moreover, if both are peak squares, then the bottom faces of the peak squares are on the same height because the pre-top square and the top square are adjacent.
    \qed
\end{proof}

Finally, Theorem~\ref{thm:existence_cover_partition} gives an explicit construction of a cover partition, called the natural cover partition.
An example of a natural cover partition is given in \cref{fig:natural_cover_partition_example}.
In Section~\ref{sec:local_cov_thm} this construction is used to inductively cover an unoccupied piece.

\begin{theorem}\label{thm:existence_cover_partition}
    There exists an explicit construction of a cover partition for each unoccupied piece~$U$, called the \textit{natural cover partition}.
\end{theorem}
\begin{proof}
    Let $S_0,\dots,S_\ell$ be the top path of $U$ and let $S_{\ell},\dots,S_{\ell+r}$ be the bottom path in reversed order.
    Consider the path $P = S_{\ell},\dots,S_{\ell+r} = S_0,\dots,S_{\ell-1}$.
    Let $S_{i_j}$ be the $j$-th vertex on $P$ such that~$\text{rf}(S_{i_j}) > \text{rf}(S_{i_j - 1})$ and~$(S_{i_j-1},S_{i_j})$ is of up type.
    This gives a sequence of squares~$S_{i_1},\dots,S_{i_s}$ on the path~$P$.
    Furthermore, the structure theorem (Theorem~\ref{thm:boundedness_structure_thm}) states that there is at most one arrow on path~$P$ of down type, this is the arrow from the end square to the penultimate square, that is, $(S_\ell,S_{\ell+1})$.

    If $(S_\ell,S_{\ell+1})$ is of down type, then consider the sequence of vertices 
    $S_{\ell}$, $S_{i_1}$, $\dots,S_{i_s}$. 
    Otherwise just consider the sequence $S_{i_1},\dots,S_{i_s}$ without the end square $S_\ell$.
    Sort $\text{bf}(S_\ell)$, $\text{bf}(S_{i_1}),\dots,\text{bf}(S_{i_s})$ by increasing height and partition~$U$ in this order.
    Inductively define the subpiece $V_j$ to be the space below $\text{bf}(S_{i_j})$ that is connected to the bottom face of $S_{i_j}$ and disjoint from the already constructed subsets of~$U$.
    For the end square $S_\ell$ denote the subpiece by $V_{\text{end}}$.
    At last, let~$V_{\text{top}}$ be the space connected to the bottom face of the top square and disjoint from $V_{\text{end}},V_1,\dots,V_{s}$.
    In the end this gives a collection of subspaces $\mathcal{P}_{\text{cov}}=\{V_1,\dots,V_s,V_{\text{end}},V_{\text{top}}\}$ (or without $V_{\text{end}}$).
    Discard the sets of measure zero.
    Call $S_{i_j}$ the \textit{square corresponding to $V_j$}.

    \textbf{Claim 1}: The collection~$\mathcal{P}_{\text{cov}}$ partitions $U$.
    To prove this claim, observe that by definition the sets are disjoint.
    Furthermore, suppose that there is a point $p$ in $U$ that is not covered by the sets in $\mathcal{P}_{\text{cov}}$.
    Then also none of the space directly above $p$ is covered by the sets, nor any space next to this vertical line.
    In other words, this uncovered space must be adjacent to a peak square.
    By Lemma~\ref{lem:peak_squares} the top square and possibly the pre-top square are the only peak squares.
    Hence the point $p$ is actually covered by $V_\text{top}$.
    Thus, it follows that the sets in $\mathcal{P}_{\text{cov}}$ cover all the space in $U$.
    Hence these sets form a partition of $U$.

    \textbf{Claim 2}: The collection~$\mathcal{P}_{\text{cov}}$ forms a cover partition of~$U$. 
    In other words, it has to be shown that the sets in $\mathcal{P}_{\text{cov}}$ are a minimal partition satisfying~(1) and~(2) from Definition~\ref{def:cover_partition}.
    Property~(1), connectedness of the sets in~$\mathcal{P}_{\text{cov}}$, follows by definition.
    For property~(2) use contradiction.
    Suppose that $V\in\mathcal{P}_{\text{cov}}$ does not satisfy property~(2), then there exists a pair of lines $\ell,\ell'\in\mathcal{L}V$ with $\ell$ below $\ell'$ such that~$\phi(\ell)\cap \phi(\ell') \neq \varnothing$ and one of the endpoints of~$\phi(\ell)$ is outside of~$\phi(\ell')$.
    Now distinguish two cases.

    \textbf{Case 1}: The left endpoint of~$\phi(\ell)$ is outside of~$\phi(\ell')$.
    Let $S$ be a square on the path~$P$ intersecting the left endpoint of~$\ell$ and let $S'$ be a square on the path~$P$ intersecting the left endpoint of~$\ell'$
    Then it holds that~$\text{rf}(S')>\text{rf}(S)$ and $S$ lies before $S'$ on the path~$P$.
    It follows that there is a vertex $S_{i_m}$ in the constructed sequence that is between $S$ and $S'$ on the path~$P$.
    Therefore, the spaces $V_{m}$ corresponding to~$S_{i_m}$ and~$V$ overlap, contradicting that~$\mathcal{P}_{\text{cov}}$ is a partition.

    \textbf{Case 2}: The right endpoint of~$\phi(\ell)$ is outside of~$\phi(\ell')$.
    Now, let $S$ be a square in the path~$P$ intersecting the right endpoint of~$\ell$ and $S'$ be a square in the path~$P$ intersecting the right endpoint of~$\ell'$.
    It holds that~$\text{lf}(S') < \text{lf}(S)$ and~$S'$ comes before~$S$ on the path~$P$.
    These two properties can only be satisfied if there is an arrow of right or down type between $S'$ and $S$ on the path~$P$.
    By the structure theorem (Theorem~\ref{thm:boundedness_structure_thm}), there does not exists a right type arrow  on the path~$P$.
    As mentioned above, the only arrow on the path~$P$ of down type is the arrow from the end square to the penultimate square.
    Now it follows that $V_\text{end}$ intersects~$V$.
    This contradicts that $\mathcal{P}_{\text{cov}}$ forms a partition.

    Finally, it remains to show that the partition is minimal.
    Let $V_j$ and $V_{j'}$ be two sets in the partition with~$j<j'$.
    Suppose that the union is connected, then~$V_{j'}$ is above $V_{j}$, as otherwise~$V_j$ could have been chosen larger in the inductive definition.
    Now the square corresponding to $V_j$ shows that~$V_j\cup V_{j'}$ cannot satisfy property~(2).
    Thus~$V_j\cup V_{j'}$ is either disconnected or does not satisfy property~(2), thus the cover partition is minimal.
    This proves Claim~2.
    The  constructed cover partition is called the natural cover partition.
    \qed
\end{proof}

In the following, it will be useful to switch between a subpiece~$V_j$ of the natural cover partition and the square~$S_{i_j}$ corresponding to this subpiece in the inductive definition of the proof of Theorem~\ref{thm:existence_cover_partition}.
\begin{definition}
    Let $\mathcal{P}_{\text{cov}}(U) = \{V_1,\dots,V_s,V_{\text{end}},V_{\text{top}}\}$ be the natural cover partition of an unoccupied piece.
    The \textit{square corresponding to} $V_j$ is defined as the square~$S_{i_j}$ in the path~$P$ such that~$\text{rf}(S_{i_j}) > \text{rf}(S_{i_j - 1})$ and~$(S_{i_j-1},S_{i_j})$ is of up type.
    The end square is the \textit{square corresponding to $V_{\text{end}}$} and the top square is the \textit{square corresponding to $V_{\text{top}}$}.
\end{definition}

The square corresponding to the green area in \cref{fig:natural_cover_partition_example} is $S_1$ and the square corresponding to the yellow area is $S_2$.
Furthermore, the end square is the square corresponding to the purple area and the top square is the square corresponding to the orange area.

\subsection{The local cover theorem}\label{sec:local_cov_thm}
The preceding sections studied the relation of squares relative to each other in the bottom-left placement of a SSPP-instance.
However, it was never actually used that all rectangles were squares, that is, the statements would also hold for SPP-instances.
On the contrary, this section will make extensive use of the fact that all rectangles are squares.
The fact that the width and height of a square are equal is important information to find a feasible covering of the unoccupied space.

This section starts by defining the width and height of an unoccupied piece and its subspaces.
Next, Lemma~\ref{lem:wide_squares} relates the size of a subpiece in the natural cover partition to the size of the square corresponding to the subpiece.
This will be the bedrock of the local cover theorem (Theorem~\ref{thm:local_cover_thm}) that covers a left or middle unoccupied piece by at most four copies of the squares that are adjacent to it.
Furthermore, Theorem~\ref{thm:right_pieces} shows that right pieces can also be covered by at most four copies of the squares adjacent to it, however, this requires more work as part of the right boundary of a right piece is the right strip boundary.
Section~\ref{sec:trenches} will discuss how to cover the remaining unoccupied space and Section~\ref{sec:global_cov_thm} unites all these local coverings into a global covering that uses at most twelve copies of the squares of the instance to cover all the unoccupied space.

The width of a subspace of an unoccupied piece is defined as the maximum width of a line in its linespace.
The height of an unoccupied piece is defined as the height of its bounding box, that is, of the smallest rectangle bounding the piece.
\begin{definition}
    Let $V$ be a subspace of an unoccupied piece $U$.
    The \textit{width} of $V$ is defined as 
    \begin{align*}
        w(V) ~=~ \max\{w(\ell)\mid \ell\in\mathcal{L}V\}.
    \end{align*}
    The \textit{height} of $V$ is defined as
    \begin{align*}
        h(V) ~=~ \sup\{y-y' \mid [x_0,x_1]\times\{y\}\in \mathcal{L}V,\ [x_0',x_1']\times\{y'\}\in \mathcal{L}V\}.
    \end{align*}
\end{definition}

For a cover partition, the fact that higher lines in a subpiece have larger width implies that the width of a subpiece is always attained in the highest line.

For each square in the bottom-left packing there is a (formal) square adjacent to the left face as otherwise the square could have been placed more to the left.
Similarly, there is a (formal) square adjacent to the bottom face, as otherwise the square could have been placed lower.
Inductively it follows that for each square there is a path in the adjacency graph consisting of right type arrows from the left strip boundary to the square.
Analoguously, there is a path consisting of up arrows from strip bottom to the square.
The next definition describes two subgraphs of the adjacency graph of an instance that contain such paths.
\begin{definition}
    The \textit{right type adjacency graph}~$G_{\text{right}}(\mathcal{I})$ is the subgraph of the adjacency graph~$G_{\text{adj}}(\mathcal{I})$ consisting of all vertices together with all right type arrows.
    Similarly, the \textit{up type adjacency graph}~$G_{\text{up}}(\mathcal{I})$ is the subgraph consisting of all up type arrows.
\end{definition}

Lemma~\ref{lem:wide_squares} relates the width of a subpiece in the natural cover partition to the width of the square corresponding to the subpiece.
Moreover, in the case that the arrow in the bottom path from the penultimate square to the end square is of up type, the lemma says even more about the width of the pretop square and top square.
In the proof it will be significant that an~$i$-piece is not yet bounded when only the first $i-1$ squares of the instance are placed.
\begin{lemma}\label{lem:wide_squares}
    Let $U$ be an $i$-piece.
    Let $\mathcal{P}_{\text{cov}}(U) = \{V_1,\dots,V_s,V_{\text{end}},V_{\text{top}}\}$ be the natural cover partition of~$U$ for the subinstance $\mathcal{I}_i$ consisting of the first $i$ squares of SSPP-instance~$\mathcal{I}$.
    Then
    \begin{enumerate}
        \item[(a)] For every square $S$  corresponding to subpiece $V \in \mathcal{P}_\text{cov}(U) \setminus \{V_\text{end}\}$, it holds that $S > w(V)$.
    \end{enumerate}
    Furthermore, if the arrow from the penultimate square to the end square is of up type, then
    \begin{enumerate}
        \item[(b)] The pretop square is larger than the height difference between the bottom face of the top square and the bottom face of the end square, that is, $S_{\text{pre}} > \text{bf}(S_{\text{top}}) - \text{bf}(S_{\text{end}})$.

        \item[(c)] The top square is larger than the penultimate square plus the width of a line in the line space of $U$ just under the bottom face of the end square, that is, $S_{\text{top}} > S_{\text{pen}} + w(V_\text{end})$.
    \end{enumerate}
\end{lemma}
\begin{proof}
    Let $S$ be the square corresponding to $V\in\mathcal{P}_{\text{cov}}(U)\setminus \{V_{\text{end}}\}$ and let $\ell=[x,x+w(V)]\times\{y\}$ be the highest line in $\mathcal{L}V$.
    Consider the space $K = [x,x+w(V)] \times [y,y+S]$ of width $w(V)$ and height $S$ above the line~$\ell$.
    Notice that $S\cap K$ is non-empty, as part of the line $\ell$ is adjacent to the bottom face of $S$.
    Assume that~$S$ is the $j$-th square that is placed.

    \textbf{Claim 1}: The interior of $K$ is unoccupied when only the first $j-1$ squares are placed.
    This claim is proven by contradiction.
    Suppose that there is a square $S'$ intersecting the interior of~$K$.
    Then there is a path $P_{\text{right}}$ from the left strip boundary to~$S'$ in the right type adjacency graph~$G_{\text{right}}(\mathcal{I}_{j-1})$ consisting of the first $j-1$ squares.
    Similarly, there is a path~$P_{\text{up}}$ from the strip bottom to~$S'$ in~$G_{\text{up}}(\mathcal{I}_{j-1})$.
    It holds that $\text{bf}(S') < \text{tf}(S)$ and $\text{lf}(S') > \text{rf}(S)$, because~$S'$ intersects the interior of~$K$.
    It follows that the path~$P_{\text{up}}$ is to the right of $S$.
    Furthermore, suppose that~$P_{\text{right}}$ goes over~$S$, then $U$ was already bounded by the paths~$P_{\text{right}}$ and~$P_{\text{up}}$ when the first~$j-1$ squares were placed, contradicting the definition of an~$i$-piece.
    Thus, the path~$P_{\text{right}}$ goes underneath~$S$.
    This means that~$P_{\text{right}}$ crosses~$V$, because~$V$ is just under~$K$ and all arrows in~$P_{\text{right}}$ are of right type, hence this contradicts that~$V$ is unoccupied.
    This proves the claim.

    Suppose that $S \leq w(V)$, then $S$ could have been placed lower as $K$ is unoccupied.
    Therefore, it holds that $S > w(V)$.
    This proves part~(a).
    
    Now, assume that the arrow from the penultimate square to the end square is of up type.
    \textbf{Claim 2}: The end square is the $i$-th square of the instance, that is, it is the square that bounds~$U$.
    The path from the end square to the left strip boundary in the right type adjacency graph has to go over~$U$, because the arrow from the penultimate square to the end square is of up type and the unoccupied piece is on the left of the bottom path.
    Thus when the end square is placed the piece is bounded, implying the claim.

    The structure theorem (Theorem~\ref{thm:boundedness_structure_thm}) implies that the arrow from the pretop square to the top square is of up type and the arrow from the top square to the end square is of right type.
    Now, suppose it holds that $S_{\text{pre}} \leq~\text{bf}(S_{\text{top}}) - \text{bf}(S_{\text{end}})$, then as the pretop square is placed before the end square, and $\text{bf}(S_{\text{top}}) -~\text{bf}(S_{\text{end}}) \leq~S_{\text{end}}$, it follows that the pretop square could have been placed lower at the position of the end square.
    This proves part~(b).

    Finally, part~(b) implies that $\text{bf}(S_{\text{end}}) \geq \text{bf}(S)$ for all squares $S\in \mathcal{P}_\text{cov} \setminus \{S_{\text{top}}\}$.
    Similar to part~(a), consider the highest horizontal line~$\ell = [x,x+w(V_{\text{end}})]\times \{y\}$ in the line space $\mathcal{L}V_{\text{end}}$ and define the space~$K = [x ,x + w(V_{\text{end}}) + S_{\text{pen}}]\times [y,y + S_{\text{top}}]$ of width $w(V_{\text{end}}) + S_{\text{pen}}$ and height $S_{\text{top}}$ above the line $\ell$. 
    Now, the interior of $K$ is unoccupied when only the squares before the top square are placed.
    This is proven with contradiction in the same way as Claim~1.
    Next, suppose that it holds that $w(S_{\text{top}}) \leq w(\ell) + w(S_{\text{pen}})$, then the top square could have been placed lower as $K$ is unoccupied.
    Hence, it holds that $w(S_{\text{top}}) > w(\ell) + w(S_{\text{pen}})$.
    This implies part~(c).
    \qed
\end{proof}

Finally, the local cover theorem (Theorem~\ref{thm:local_cover_thm}) constructs a cover of a left or middle unoccupied~$i$-piece using at most four copies of the squares adjacent to the piece.
The main idea is to cover each subpiece in the natural cover partition by at most two copies of the square~$S$ corresponding to the subpiece, this is possible as the width of the subpiece is smaller than~$S$.
However, the height of the subpiece might be larger than~$S$, in that case, the other square adjacent to the subpiece on the same height as~$S$ is also used to cover the subpiece.
\begin{theorem}[Local cover theorem]\label{thm:local_cover_thm}
    Let $U$ be a left or middle $i$-piece in the bottom-left packing of the subinstance~$\mathcal{I}_i$.
    Then~$U$ can be covered by at most four copies of the squares that are adjacent to~$U$.
\end{theorem}
\begin{proof}
    Let $\mathcal{P}_{\text{cov}}(U) = \{V_1,\dots,V_s,V_{\text{end}},V_{\text{top}}\}$ be the natural cover partition of~$U$.
    Distinguish two cases depending on the type of the arrow from the penultimate square to the end square.

    \textbf{Case 1}: Let the arrow from the penultimate square to the end square be of right type.
    Then let $S_{i_1},\dots,S_{i_s}$ be the squares corresponding to $V_1,\dots,V_s$ and let $S_{\text{top}}$ be the top square corresponding to~$V_{\text{top}}$.
    Observe that $V_\text{end}$ does not exist.
    For~$1\leq j\leq s$, let $Q_j$ be the path from~$S_{i_j-1}$ to~$S_{\text{end}}$ that traverses the top path in reversed order followed by the bottom path.
    And define $S_{k_j}$ to be the first vertex on the path~$Q_j$ such that~$\text{tf}(S_{k_j}) > \text{bf}(S_{i_j})$.
    Obviously,~$S_{k_j}$ is adjacent to $V_j$ as otherwise $V_j$ could have been chosen to be larger.
    Now split $V_j$ into $V_j^{\bot}$ and~$V_j^{\top}$, where $V_j^{\bot}$ is everything below the bottom face of $S_{k_j}$ and $V_j^\top$ everything above the bottom face of $S_{k_j}$, that is, 
    \begin{align*}
        V_j^\bot = V_j \cap ([0,W]\times[0,\text{bf}(S_{k_j})]) \quad\text{and}\quad V_j^\top = V_j \cap ([0,W]\times [\text{bf}(S_{k_j}),\infty)).
    \end{align*}
    The structure theorem (Theorem~\ref{thm:boundedness_structure_thm}) states that all arrows in the path $Q_j$ are either of right or down type.
    Thus for each square on~$Q_j$, the effect it has on the height of~$V_j$ is less than the effect it has on the width of~$V_j$.
    Therefore it follows that $w(V_j^\bot)>h(V_j^\top)$.
    Now by part~(a) of Lemma~\ref{lem:wide_squares} it holds that $S_{i_j}>w(V_j)$, hence $S_{i_j}$ can cover $V_j^\top$.
    Furthermore, if $S_{i_j}\geq S_{k_j}$, then $S_{i_j}$ can also cover $V_j^\top$.
    Otherwise if~$S_{i_j}<S_{k_j}$, then~$S_{k_j}$ can cover $V_j^\top$.
    This is illustrated in \cref{fig:local_cover_thm}.

    \begin{figure}[ht]
        \centering
        \begin{subfigure}{0.45\textwidth}
        \begin{tikzpicture}[scale=0.53]
            \filldraw[myfill] (-3,-3) rectangle (3,3);
            \filldraw[myfill] (0,3) rectangle (4,7);
            \filldraw[myfill] (6,1) rectangle (9,4);
            \draw [dashed] (4,3) -- (6,3);
            \draw [dashed] (3,1) -- (6,1);
    
            \node[] at (2,5) {$S_{i_j}$};
            \node[] at (7.5,2.5) {$S_{k_j}$};
            \node[] at (4.5,2) {$V_{j}^\top$};
            \node[] at (4.5,0) {$V_{j}^\bot$};  
        \end{tikzpicture}~~~~
        \caption{}
        \label{fig:local_cover_thm_top_path_a}
        \end{subfigure}
        \begin{subfigure}{0.45\textwidth}
        \begin{tikzpicture}[scale=0.53]
            \filldraw[myfill] (-3,-3) rectangle (3,3);
            \filldraw[myfill] (0,3) rectangle (4,7);
            \filldraw[myfill] (6,-1) rectangle (11,4);
            \draw [dashed] (4,3) -- (6,3);
            \draw [dashed] (3,-1) -- (6,-1);
    
            \node[] at (2,5) {$S_{i_j}$};
            \node[] at (8.5,1.5) {$S_{k_j}$};
            \node[] at (4.5,1) {$V_{j}^\top$};
            \node[] at (4.5,-2) {$V_{j}^\bot$};  
        \end{tikzpicture}
        \caption{}
        \label{fig:local_cover_thm_top_path_b}
        \end{subfigure}
        \caption{In Figure~\ref{fig:local_cover_thm_top_path_a} the square corresponding to $S_{i_j}$ is larger than the square corresponding to $S_{k_j}$, so $S_{i_j}$ is used twice to cover $V_j^\top$ and $V_j^\bot$.
        Furthermore, in Figure~\ref{fig:local_cover_thm_top_path_b} the square corresponding to $S_{i_j}$ is smaller than the square corresponding to $S_{k_j}$, so $S_{i_j}$ is used once to cover $V_j^\bot$ and the area in $S_{k_j}$ directly next to~$V_j^\top$ is used once to cover $V_j^\top$.}
        \label{fig:local_cover_thm}
    \end{figure}
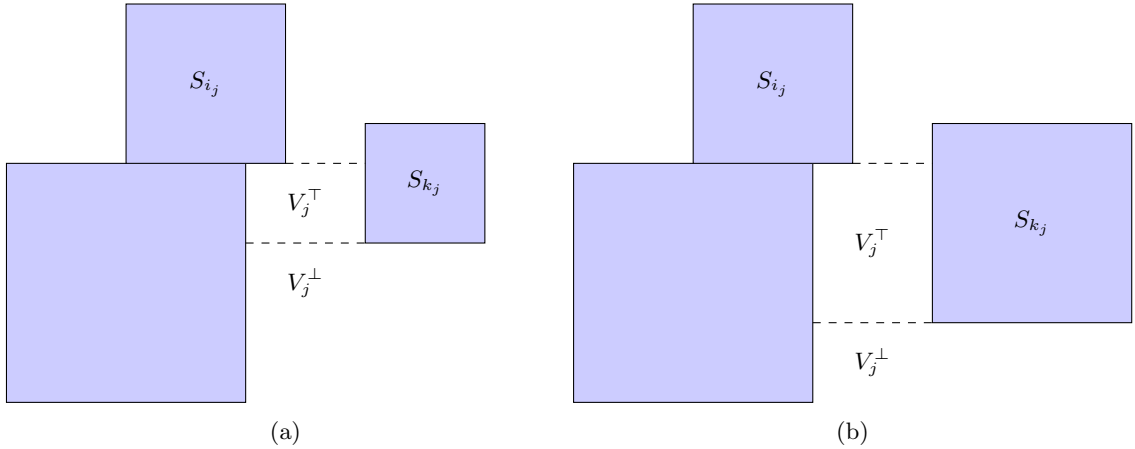

    In the same way, let $V_{\text{top}}^\bot$ be all the space in~$V_{\text{top}}$ under $\text{bf}(S_{\text{end}})$ and  $V_{\text{top}}^\top$ all the space in~$V_{\text{top}}$ that is above~$\text{bf}(S_{\text{end}})$. 
    Again, part~(a) of Lemma~\ref{lem:wide_squares} states that $S_{\text{top}} > w(V_{\text{top}})$, thus either two copies of $S_{\text{top}}$ or one copy of $S_{\text{top}}$ and one copy of $S_{\text{end}}$ cover $V_{\text{top}}$.

    All in all, at most three copies of the squares adjacent to~$U$ cover the unoccupied piece.
    Namely, for each subpiece use at most two copies of~$S_{i_j}$ and at most one copy of~$S_{k_j}$.
    Observe that if $S_{k_j}$ is used to cover $V_j^\top$, then only $S_{k_j}\cap([0,W]\times [\text{bf}(V_j^\top),\text{tf}(V_j^\bot)])$ is needed, thus $S_{k_j}$ can be used to cover multiple subpieces.
    Thus each square might be used at most three times in the covering of the piece~$U$.

    \textbf{Case 2}: Let the arrow from the penultimate square to the end square be of up type.
    Now, let $S_{i_1},\dots,S_{i_s}$ be the squares corresponding to $V_1,\dots,V_s$, let $S_{\text{top}}$ the square corresponding to~$V_{\text{top}}$ and $S_{\text{end}}$ the square corresponding to $V_{\text{end}}$.
    Cover $V_{\text{top}}$ and $V_j$ for $1\leq j\leq s$ in the same way as Case~1.
    
    Next, the subpiece $V_{\text{end}}$ is covered by at most two copies of the top square.
    Again split~$V_{\text{end}}$ into $V_{\text{end}}^\bot$ and $V_{\text{end}}^\top$, where $V_{\text{end}}^\bot$ is everything in $V_{\text{end}}$ that is under~$\text{bf}(S_{\text{pen}})$, and $V_{\text{end}}^\top$ everything above~$\text{bf}(S_{\text{pen}})$.
    By part~(c) of Lemma~\ref{lem:wide_squares} it holds that~$S_{\text{top}}>S_{\text{pen}} + w(V_{\text{end}})$, this implies that $S_{\text{top}} > S_{\text{pen}}$ and $S_{\text{top}}>w(V_{\text{end}})$.
    Now by the same argumentation as before, the structure theorem (Theorem~\ref{thm:boundedness_structure_thm}) states that all arrows on the bottom path are of right or up type, therefore, each square in the bottom path adjacent to~$V_{\text{end}}^\bot$ adds more to the width than to the height, thus $w(V_{\text{end}}^\bot)>h(V_{\text{end}}^\bot)$.
    This implies that the top square can cover~$V_{\text{end}}^\bot$.
    Furthermore, $S_{\text{top}} > S_{\text{pen}}$ implies that the top square can cover~$V_{\text{end}}^\top$.

    All in all, at most four copies of the squares adjacent to $U$ cover the unoccupied piece.
    Each subpiece~$V_j$ uses at most two copies of $S_{i_j}$.
    At most one copy is needed of a square that plays the role of some~$S_{k_j}$.
    And the top square is used at most four times, namely, at most twice to cover~$V_{\text{top}}$ and at most twice to cover~$V_{\text{end}}$.
    As a top square is never used in the role of~$S_{k_j}$, it follows that at most four copies of the squares adjacent to~$U$ suffices to cover the piece.
    \qed
\end{proof}

Later, the global cover theorem (Theorem~\ref{thm:global_cover_thm}) will construct a global covering of the unoccupied space using the local cover theorem (Theorem~\ref{thm:local_cover_thm}) multiple times.
It will be crucial that each square is only used a limited number of times to cover different local coverings.
As mentioned in the end of the previous proof, a square can be used to cover unoccupied space in two different roles.
First of all, a square can be used to cover a subpiece~$V_j$ in the role of the square~$S_{i_j}$ corresponding to~$V_j$.
Secondly, a square can be used to cover~$V_j^\top$ in the role of the square~$S_{k_j}$, which is the first square on the path~$Q_j$ such that the top face of~$S_{k_j}$ is strictly above the bottom face of $S_{i_j}$.
The next definition makes this distinction formal.
\begin{definition}
    A square~$S$ is a \textit{left cover square} if $S$ is used to cover a subpiece~$V_j$ of an unoccupied piece~$U$ in the role of the square~$S_{i_j}$ corresponding to~$V_j$.
    A square is a \textit{right cover square} if $S$ is used to cover a subpiece~$V_j$ in the role of the square~$S_{k_j}$, which is the first square on the path~$Q_j$ between the square $S_{i_j-1}$ and the end square containing the start square such that $\text{tf}(S_{k_j}) > \text{bf}(S_{i_j})$.
\end{definition}

The local cover theorem (Theorem~\ref{thm:local_cover_thm}) gives a recipe for covering left and middle pieces, thus it remains to construct a covering for right pieces.
The covering from before does not work, because for a subpiece~$V_j$ the right cover square $S_{k_j}$ on the path~$Q_j$ might be the formal right strip boundary square and this cannot be used to cover~$V_j$.
Despite of this, Theorem~\ref{thm:right_pieces} constructs a cover that heavily relies on the local cover theorem (Theorem~\ref{thm:local_cover_thm}).
Namely, it uses the local cover theorem for subpieces that have a right cover square adjacent to it, and it uses other squares in the top path to cover the other unoccupied space.

\begin{theorem}[Right piece local cover theorem]\label{thm:right_pieces}
    Let $U$ be a right $i$-piece in the bottom-left packing of the subinstance~$\mathcal{I}_i$.
    Then $U$ can be covered by at most four copies of the squares adjacent to~$U$.
\end{theorem}
\begin{proof}
    Let $\mathcal{P}_{\text{cov}}(U) = \{V_1,\dots,V_s,V_{\text{top}}\}$ be the natural cover partition of~$U$. 
    Note that $V_{\text{end}}$ does not exist for a right piece.
    For each subpiece~$V_j$ such that $S_{k_j}$ does not equal the formal right strip boundary, use the same covering with~$S_{i_j}$ and~$S_{k_j}$ as in Theorem~\ref{thm:local_cover_thm}.
    Now let~$V_{l_1},\dots,V_{l_t}$ be the other subpieces that are not yet covered (including $V_{\text{top}}$).

    For each~$1\leq j \leq t$ it holds that $S_{i_{l_j}} > w(V_{l_j})$ by part~(a) of Lemma~\ref{lem:wide_squares}.
    However, the height of $V_{l_j}$ can be significantly larger than~$S_{i_{l_j}}$, because there is not a bottom path consisting of right and down type arrows on the right of the subpiece.
    Instead, let $\ell$ be the bottom-left most horizontal line in $\mathcal{L}V_{i_l}$ and let $S$ be a square adjacent to the left endpoint of $\ell$.
    Consider the path $w_1,\dots,w_q$ from $S$ to $S_{i_{l_j}}$.
    By the structure theorem (Theorem~\ref{thm:boundedness_structure_thm}) this path consists of left and up type arrows.
    Next the piece~$V_j$ is partitioned into $q$ classes $W_1,\dots,W_q$ where~$W_k$ is the unoccupied space in the rectangle $[\text{rf}(w_k),W]\times [\text{tf}(w_{k-1}),\text{tf}(w_k)]$ that is connected to the square~$w_k$.
    These sets form a partition, because if there is other unoccupied space in the rectangle corresponding to~$W_k$, then there is another square~$w_{k'}$ on the top path such that right of that square there is another set~$W_{k'}$ of the partition.
    Each $W_k$ is covered by either $w_k$, $w_{k+1}$ or $S_{i_j}$ depending on the arrow types of $(w_{k-1},w_k)$ and $(w_k,w_{k-1})$.

    \textbf{Case 1}: Let $(w_{k-1},w_k)$ and $(w_k,w_{k-1})$ be of left type.
    Then the effect of~$w_k$ on the height of~$V_j$ is strictly less than the effect on the width of~$V_j$.
    Thus cover $W_k$ with $S_{i_j}$.
    
    \textbf{Case 2}: Let $(w_{k-1},w_k)$ and $(w_k,w_{k-1})$ be of up type. 
    Then it holds that $w(W_k) < w_{k+1}$, because otherwise~$w_{k+1}$ could have been placed lower.
    Now if $w_k < w_{k+1}$, then as $w_k = h(W_k)$ it is possible to cover~$W_k$ with $w_{k+1}$.
    Otherwise if $w_k \geq w_{k+1}$, then it holds that $w_k>w(W_k)$, thus $W_k$ can be covered with~$w_k$.
    
    \textbf{Case 3}: Let $(w_{k-1},w_k)$ be of left type and $(w_k,w_{k-1})$ of up type.
    This is similar to Case~2.
    If $w_k < w_{k+1}$, then cover $W_k$ with $w_{k+1}$.
    Otherwise cover~$W_k$ with $w_k$.
    
    \textbf{Case 4}: Let $(w_{k-1},w_k)$ be of up type and $(w_k,w_{k-1})$ of left type.
    Then the effect of $w_k$ on the height of~$V_j$ equals the effect on the width of $V_j$.
    Thus cover $W_k$ with $S_{i_j}$.

    Let $W_{m_1},\dots,W_{m_r}$ be the sets from the partition that were covered with~$S_{i_j}$ in Case~1 and~4.
    As mentioned above, for each $W_k$ it holds that $S_{i_j}>w(W_k)$.
    Furthermore, as the effect of $w_{m_k}$ on the height of $V_j$ is less or equal to the effect on the width of~$V_j$ it follows that~$\sum_{x=1}^r w_{m_x} \geq \sum_{x=1}^r h(W_{m_x})$.
    Hence it holds that
    \begin{align*}
        S_{i_j} ~>~ w(V_j) ~\geq~ \sum_{x=1}^r w_{m_x} ~>~ \sum_{x=1}^r h(W_{m_x}).
    \end{align*}
    Thus one copy of $S_{i_j}$ is enough to cover $W_{m_1},\dots,W_{m_r}$.

    All in all, a square is used at most three times to cover subpieces of the natural cover partition by the local cover theorem (Theorem~\ref{thm:local_cover_thm}), because the arrow from the penultimate square to the end square is of right type.
    Furthermore, a square can be in at most two of the four cases above.
    Hence at most two copies of a square are needed to cover the $W_1,\dots,W_q$, namely, at most once in Case~$2$, at most once in Case~$3$ and at most once in Case~$1$ and~$4$.
    Moreover, a square that is used more than twice to cover subpieces in the local cover theorem can only belong to Case~$1$ above (or to none of the cases).
    This is true because such a square is a left and right cover square, hence there is unoccupied space in~$U$ adjacent to the left face and to the right face of the square.
    Therefore, it follows that the arrow to and from this square in the top path must be of left type by the construction in Lemma~\ref{lem:hamcircuit}.
    Therefore it follows that in total at most four copies of the squares adjacent to the right piece are needed to cover it.
    \qed
\end{proof}

\subsection{Trenches}\label{sec:trenches}
The unoccupied space of a bottom-left packing is usually larger than just the unoccupied pieces.
In particular, there can be space in the top of the packing that is unbounded in the strip.
These subspaces are called trenches.
Informally, a trench looks like an unoccupied piece without a top square.
Theorem~\ref{thm:top_trenches} and Theorem~\ref{thm:right_trench} reduce trenches to left, middle or right pieces to show that part of a trench can be covered by at most four copies of squares that are adjacent to it.
This will be used in Section~\ref{sec:global_cov_thm} to construct a global covering of all the unoccupied space of the packing.
\begin{definition}
    Let $\mathcal{I}$ be a SSPP-instance and let $\mathcal{U}$ be the set of all unoccupied pieces.
    A~\textit{trench} is a bounded connected maximal subspace $T$ of
    \begin{align*}
        ([0,W]\times[0,h_{\text{BL}}-h_{\text{max}}]) \setminus (\text{BL}(\mathcal{I}) \cup \mathcal{U}).
    \end{align*}
    Here $h_\text{max}$ is the height of the largest square.
    Let $\mathcal{T} = \{T_1,\dots,T_s\}$ be the set of trenches.
\end{definition}

All trenches have the same properties except for one special trench called the right trench.
The right trench is the trench that is adjacent to the right strip boundary.
In the reduction theorems, the right trench is reduced to a right piece instead of a left or middle piece.
\begin{definition}
    The \textit{right trench} is the unique trench that is adjacent to the right strip boundary.
    The other trenches are called \textit{top trenches}.
\end{definition}

The definition of trenches directly implies the following corollary describing the unoccupied space in a bottom-left packing in terms of unoccupied pieces and trenches.
\begin{corollary}
   \label{cor:all_unoccupied_spaces}
    The unoccupied space of a bottom-left packing in the $[0,W]\times [0,h_{\text{BL}}-h_{\text{max}}]$ substrip is contained in the union of the unoccupied pieces and the trenches, that is~$\mathcal{U}\cup \mathcal{T}$.
\end{corollary}

Next, Theorem~\ref{thm:top_trenches} shows that top trenches can be covered by at most four copies of the squares adjacent to it when a substrip of height $3h_{\text{max}}$ is cut off the top of the strip.
Section~\ref{sec:boundedness} will show that cutting of such a substrip does not effect the boundedness of the approximation ratio.
The main idea for covering the top trenches is to first reduce the top trenches to left and middle pieces, and then use the local cover theorem (Theorem~\ref{thm:local_cover_thm}) to cover the unoccupied space with squares adjacent to the trench.
\begin{theorem}[Top trench reduction theorem]\label{thm:top_trenches}
    Let $T$ be a top trench.
    Then the space
    \begin{align*}
        T\cap ([0,W]\times [0,h_{\text{BL}}-3h_{\text{max}}])
    \end{align*}
    can be covered by at most four copies of the squares adjacent to~$T$.
\end{theorem}
\begin{proof}
    Let $\ell$ be the highest horizontal line in the line space of~$T$.
    There is a square above the line~$\ell$, since at least one square touches the top of the strip at height $h_{\text{BL}}$, hence it holds that~$w(\ell) < h_{\text{max}}$.
    Place a square~$S$ of size~$h_{\text{max}}$ on top of the line~$\ell$ to reduce the top trench~$T$ to a left or middle piece.
    Use the local cover theorem (Theorem~\ref{thm:local_cover_thm}) to cover the piece with at most four copies of the squares adjacent to~$T$.
    The square~$S$ of size~$h_{\text{max}}$ is the top square of the unoccupied piece~$T$.
    Now~$S$ is used at most four times to cover~$T$, and~$S$ is only used to cover unoccupied space that is at most $2h_{\text{max}}$ under the line~$\ell$.
    Therefore, all the space in~$T\cap ([0,W]\times [0,h_{\text{BL}}-3h_{\text{max}}])$ is covered by actual squares that are adjacent to~$T$.
    \qed
\end{proof}

It remains to cover the right trench.
This is done similar to top trenches.
However, the difference is that the right trench is reduced to a right piece by adding a copy of the largest square of the instance to the top of the trench.
Next, Theorem~\ref{thm:right_pieces} shows a way to cover the right piece by at most four copies of the squares adjacent to it.
\begin{theorem}[Right trench reduction theorem]\label{thm:right_trench}
    The right trench~$T$ can be covered by at most four copies of the squares adjacent to $T$ together with a square of size~$h_{\text{max}}$.
\end{theorem}
\begin{proof}
    Let $\ell$ be the highest horizontal line in the line space of~$T$.
    There is a square above the line~$\ell$, hence~$w(\ell) < h_{\text{max}}$.
    Place a square~$S$ of size~$h_{\text{max}}$ on top of the line~$\ell$ to reduce the right trench~$T$ to a right piece.
    Use Theorem~\ref{thm:right_pieces} to cover $T$ by at most four copies of the squares adjacent to the right piece~$T$.
    It follows that this space can be covered by at most four copies of the squares that are actual adjacent to the right trench~$T$ together with a square of size~$h_{\text{max}}$.
    \qed
\end{proof}

\subsection{The global cover theorem}\label{sec:global_cov_thm}
The global cover theorem (Theorem~\ref{thm:global_cover_thm}) combines the different reductions with the local cover theorems to obtain a global covering of the unoccupied space in a bottom-left packing of squares restricted to the $[0,W]\times [0,h_{\text{BL}}- 3h_{\text{max}}]$ substrip.
The main caveat is to show that each square is only used a limited number of times to cover unoccupied space locally.
Essentially, the reductions from the trenches to the unoccupied pieces show that each square covers unoccupied space as left cover square, as right cover square, or it covers a right piece.
The idea of the global cover theorem is to show that if a square is used multiple times as left or right cover square, then still only a limited number of copies of this square is required.
\begin{theorem}[Global cover theorem]\label{thm:global_cover_thm}
    At most twelve copies of the squares from a SSPP-instance are needed to cover the $[0,W]\times [0,h_{\text{BL}}- 3h_{\text{max}}]$ substrip.
\end{theorem}
\begin{proof}
    By Corollary~\ref{cor:all_unoccupied_spaces} all the unoccupied space of the substrip $[0,W]\times [0,h_{\text{BL}}- 3h_{\text{max}}]$ is contained in the union of the pieces and trenches.
    Use the local cover theorem (Theorem~\ref{thm:local_cover_thm}) to cover the left and middle unoccupied pieces, use Theorem~\ref{thm:right_pieces} to cover the right pieces, use Theorem~\ref{thm:top_trenches} to cover the top trenches and use Theorem~\ref{thm:right_pieces} to cover the right trench.
    It remains to show that each square is only used a limited number of times to cover unoccupied space.

    \textbf{Claim 1}: If a square~$S$ is used multiple times as left cover square, then still at most six copies of the square are needed.
    Let $U_1,\dots,U_l$ be the different pieces that use $S$ as left cover square in their local covering.
    Then $S$ is the top square of each $U_j$ except possibly for one such piece.
    Without loss of generality, let $S$ be the top square of the pieces $U_1,\dots,U_{l-1}$.
    Let~$V_1,\dots,V_{l-1}$ be the subpieces of the natural cover partition of $U_1,\dots,U_{l-1}$ that use~$S$ in their local covering.
    Obviously, it holds that $\sum_{j=1}^{l-1} w(V_j) < S$ as~$S$ is the top square of all these pieces.
    Furthermore, the top faces of the end squares of $U_1,\dots,U_{l-1}$ are on the same height and the arrow from the top square~$S$ of $U_j$ to the end square of~$U_j$ is of down type.
    This implies that the arrow from the penultimate square to the end square of~$U_j$ for $1\leq j\leq l-1$ is of right type.
    Hence, to cover~$V_j$ for $1\leq j\leq l-1$ at most two copies of the space in $S$ directly above~$V_j$ are needed.
    Furthermore, at most four copies of~$S$ are required to cover~$V_l$ according to the local cover theorem (Theorem~\ref{thm:local_cover_thm}).
    Thus in total at most six copies of~$S$ are needed for~$S$ in the role of left cover square to cover the unoccupied space.

    \textbf{Claim 2}: If a square is used multiple times as right cover square, then at most one copy of the square is needed.
    To prove this claim, let $V_1,\dots,V_l$ be the different subpieces that use $S$ as right cover square in their local covering.
    These subpieces might belong to different pieces.
    For each~$j$, the square~$S$ is used to cover $V_j^\top$ as defined in the proof of the local cover theorem (Theorem~\ref{thm:local_cover_thm}).
    Now it holds that $\sum_{j=1}^l h(V_j^\top) < S$, because each point on the right boundary of $V_j^\top$ is adjacent to~$S$.
    Furthermore, for each~$j$ it holds that $w(V_j^\top) < S$, thus in total at most one copy of~$S$ is needed for $S$ in the role of right cover square to cover the unoccupied space.

    \textbf{Claim 3}: Each square is needed at most twelve times to cover unoccupied space.
    A square is used at most six times as left cover square and at most one time as right cover square.
    Furthermore, to cover the right trench an extra copy of the largest square is needed.
    Last of all, a square never has to cover more than one right piece, and this requires at most four copies of the square.
    Thus $6+1+1+4 = 12$ copies of the squares suffices to cover all the unoccupied space in the bottom-left packing restricted to the $[0,W]\times [0,h_{\text{BL}}- 3h_{\text{max}}]$ substrip.
    \qed
\end{proof}

\subsection{Boundedness for SSPP}\label{sec:boundedness}
We now show that the approximation ratio of the bottom-left algorithm for any ordering of squares is bounded by a constant.
Therefore, even when the worst ordering of the squares is used, the approximation ratio remains bounded.

Although it might be interesting to get an approximation ratio as small as possible, this section only cares about boundedness.
Theorem~\ref{thm:worst_case_bound}  expresses the bound in terms of abstract numbers, illustrating that improvement can be found by lowering any of the constants involved.
Most certainly improvement is possible by either constructing a different covering or by enhanching the analysis of this section.
Next, Corollary~\ref{cor:worst_case_bound} substitutes the numbers found in the global cover theorem (Theorem~\ref{thm:global_cover_thm}) to obtain a $16$-approximation.
\begin{theorem}\label{thm:worst_case_bound}
    Let $\mathcal{I}$ be a Square Strip Packing instance, then 
    \begin{align*}
        \frac{h_{\text{BL}}^{\text{worst}}(\mathcal{I})}{h_{\text{OPT}}(\mathcal{I})} ~\leq~ f+g+1.
    \end{align*}
    Here $f$ is the number of copies of the squares in the instance that is required to cover the unoccupied space in the $[0,W] \times [0,h_{\text{BL}}-gh_{\text{max}}]$ substrip for some constant $g$.
\end{theorem}
\begin{proof}
    The total area of the strip~$[0,W]\times[0,h_{\text{BL}}]$ equals the total area of the squares~$A_{\text{squares}}$ plus the unoccupied area~$A_{\text{unocc}}$.
    The unoccupied area is bounded by $f$ copies of the squares plus the area outside of the $[0,W] \times [0,h_{\text{BL}} - gh_{\text{max}}]$ substrip.
    In other words, it holds that
    \begin{align*}
        A_{\text{unocc}} ~\leq~ f A_{\text{squares}} + g h_{\text{max}} W.
    \end{align*}
    Furthermore, the total area of the squares is bounded by the total area of an optimum packing, therefore it holds that
    \begin{align*}
        h_{\text{BL}} W ~=~ A_{\text{squares}} + A_{\text{unocc}} ~\leq~ (f+1) A_{\text{squares}} + g h_{\text{max}} W ~\leq~ (f+1) h_{\text{OPT}} W + g h_{\text{max}} W.
    \end{align*}
    It holds that $h_{\text{max}}\leq h_{\text{OPT}}$.
    Thus this implies that
    \begin{align*}
        h_{\text{BL}} ~\leq~ (f+1) h_{\text{OPT}} + g h_{\text{max}} ~\leq~ (f+g+1) h_{\text{OPT}}.
    \end{align*}
    \qed
\end{proof}

The analysis of the previous proof actually gives an asymptotic approximation ratio, namely it shows that~$h_{\text{BL}} \leq (f+1)h_{\text{OPT}} + g h_{\text{max}}$.
However, this is irrelevant for boundedness.
Next, Corollary~\ref{cor:worst_case_bound} uses the global cover theorem (Theorem~\ref{thm:global_cover_thm}) and Theorem~\ref{thm:worst_case_bound} to show that the approximation ratio of an SSPP-instance using any order is bounded by~$16$.
\begin{corollary}
    \label{cor:worst_case_bound}
    The bottom-left algorithm has constant approximation ratio for the Square Strip Packing Problem, it holds that $\hblworst(\mathcal{I}) \leq 16 \cdot\hopt(\mathcal{I})$.
\end{corollary}
\begin{proof}
    According to the global cover theorem (Theorem~\ref{thm:global_cover_thm}), at most twelve copies of the squares from the instance~$\mathcal{I}$ are needed to cover the unoccupied space in the $[0,W]\times [0,h_{\text{BL}}-3h_{\text{max}}]$ substrip.
    Hence, Theorem~\ref{thm:worst_case_bound} implies that
    \begin{align*}
        \frac{h_{\text{BL}}^{\text{worst}}(\mathcal{I})}{h_{\text{OPT}}(\mathcal{I})} ~\leq~ f+g+1 ~=~ 12 + 3 + 1 ~=~ 16.
    \end{align*}
    \qed
\end{proof}

\end{document}